\documentclass{astlb}

\usepackage{graphicx}
\usepackage{amsmath}
\usepackage{amssymb}
\usepackage{mathtools}
\usepackage{gensymb}

\usepackage{array}
\usepackage{tabularx}
\usepackage{multirow}
\usepackage[hyperfootnotes=false]{hyperref}

\newcolumntype{C}{>{\small\centering\arraybackslash}X}

\def\degree{^{\circle}}

\def\msun{M_{\odot}}

\def\a0535{1A~0535+26}
\def\ks1947{KS~1947+300}
\def\g1008{GRO~J1008-57}
\def\v0332{V~0332+53}
\def\2s1553{2S~1553-542}
\def\gro1750{GRO~J1750-27}
\def\4u0115{4U~0115+634}
\def\xte1946{XTE~J1946+274}
\def\rx0520{RX~J0520.5-6932}
\def\exo2030{EXO~2030+375}

\begin{document}

\journalinfo{2017}{43}{11}{706}[729]

\title{Radius of the Neutron Star Magnetosphere during Disk Accretion}

\author{\bf E.V. Filippova\email{kate@iki.rssi.ru}\address{1}, I.A. Mereminskiy\address{1}, A.A. Lutovinov\address{1}, S.V. Molkov\address{1} S.S.Tsygankov\address{2,1}
\addresstext{1}{Space Research Institute, Moscow, Russia}
\addresstext{2}{Tuorla observatory, University of Turku, Finland}} \shortauthor{Filippova et al.} \shorttitle{ Radius of the neutron star magnetosphere during disk accretion} 
\submitted{29.11.2016}

\begin{abstract}
The dependence of the spin frequency derivative $\dot{\nu}$ of accreting neutron stars with a strong magnetic field (X-ray pulsars) on the mass accretion rate (bolometric luminosity, $L_{bol}$) has been investigated for eight transient pulsars in binary systems with Be stars. Using data from the Fermi/GBM and Swift/BAT telescopes, we have shown that for seven of the eight systems the dependence $\dot{\nu}$ can be fitted by
the model of angular momentum transfer through an accretion disk, which predicts the relation $\dot{\nu}\sim L^{6/7}_{bol}$.
Hysteresis in the dependence $\dot{\nu}(L_{bol})$ has been confirmed in the system \v0332 and has been detected for the first time in the systems \ks1947, \g1008, and \a0535. The radius of the neutron star magnetosphere in all of the investigated systems have been estimated. We show that this quantity varies from pulsar to pulsar and depends strongly on the analytical model and the estimates for the neutron star and binary system parameters.


\end{abstract}

\section*{INTRODUCTION}
X-ray pulsars are a natural laboratory for investigating the disk accretion of matter, because the structure of the inner disk region and its interaction with the neutron star magnetosphere affect the pulsar spin period. It follows from observations that the pulsation periods are subjected to variabilities in a wide range of characteristic time scales, from hours to years, due to changes in the regime and rate of accretion onto the neutron star \citep{nagase89,bildsten97,lutovinov94,lutovinov09}. One of the tools for investigating the accretion regime is to analyze the dependence of the neutron star spin frequency
derivative on the mass accretion rate, $\dot{\nu}(\dot{M})$. The accretion regime is determined by the radius of the neutron star magnetosphere that is specified as $r_m=\xi r_A$,  where $r_A=\Big({{\mu^4}\over{2GM \dot{M}^2}}\Big)^{1/7}$ is the Alfven radius determined by the equality of the magnetic pressure and the pressure of accreting matter falling spherically symmetrical from infinity to a dipole, $\xi$ is a dimensionless parameter. If the magnetospheric radius of the star is larger than its corotation radius, then the system is in the "propeller" regime \citep{illarionov75}, where the neutron star flings away the accreting matter, thereby losing its angular momentum and spinning down. The propeller effect is quite difficult to detect, 
and by now this has been done only for a few X-ray pulsars \citep{campana08,tsygankov16a,tsygankov16m82,lutovinov17smc}. If the magnetospheric radius is smaller than the corotation radius, then the neutron star, as a rule, spins up through the angular momentum accretion of the infalling matter. There exist several models that describe the change in the spin frequency of the neutron stars due to the interaction of their magnetosphere with the accretion disk. This problem has been intensively studied since the early 1970s, but the first complete solution was published in \cite{ghosh79}. These authors considered a model where the magnetic field lines of the neutron star thread a standard accretion disk. Two regions are formed in this case: the interaction must cause the neutron star to spin up at radii smaller than the corotation radius, because the Keplerian angular velocity of the disk is larger than the stellar rotation rate, and to spin down at larger ones. The dependence of the neutron star spin frequency derivative on the accretion rate in this model looks like $\dot{\nu} \sim \dot{M}^{\beta}$ , where $\beta = 6/7$.

If a magnetic diffusion in the disk is not large enough for the field lines to slide in the azimuthal direction, then the magnetic field lines connecting the star and the disk will twist as a result of different rotation rates of their bases. This will give rise to a toroidal field whose pressure will push away the field lines in such a way that they will become almost radially directed. In this case, the field lines of opposite directions can be separated by a current sheet where their reconnection can occur \citep{aly85,lynden-bell94,uzdensky02}. \cite{lovelace95} proposed a model where the field lines became open as a result of their interaction with the disk and formed a stable configuration without undergoing reconnection, which allowed the wind to take away some angular momentum from both the star and the disk. In the most recent paper \cite{parfrey16} considered a model that results from the relaxation of the model described in  \cite{lovelace95}, where the field lines passing through the accretion disk are pushed beyond the light cylinder as a result of the magnetic diffusion.

The influence of conditions in the inner disk on the dependence $\dot{\nu}(\dot{M})$ was considered by \cite{ghosh92}, who showed that (1) $\beta=0.925$ for a radiation-dominated disk, (2) $\beta=0.15$ for a two-temperature, optically thin in the vertical direction gas-dominated disk in the case of electron cooling through Comptonization of soft photons from the external source, or (3)$\beta=0.76$ in the case of electron cooling through bremsstrahlung.

It follows from analytical calculations that during accretion through a standard disk $\xi$ can be as $\sim 0.5$ \citep{ghosh77,kluzniak07} as $\gtrsim1$ \citep{wang96}. Numerical calculations show that the relative size of the neutron star magnetosphere is closer to $\sim0.5$ \citep{long05,bessolaz08,zanni13}. Recently, \cite{chashkina17} have shown that for a
radiation-dominated disk the predicted value of $\xi$ is larger than that for a standard disk by a factor of 1.5-2, with the absolute value of the magnetospheric radius being independent of the mass accretion rate and the exponent $\beta=1$. Thus, not only $\beta$ but also the magnetospheric size depends on the accretion rate in other disk states. Changes in both parameters affect the evolution of the neutron star spin frequency derivative. 

Be systems are suitable objects for investigating
the variability of the accreting neutron stars rotation rate, because they exhibit a fairly regular outburst activity, when the accretion rate changes by several orders of magnitude. These systems demonstrate two types of outbursts. The type I outbursts are periodic, coincident with the time of periastron passage, and are associated with an enhancement of
the mass accretion rate as the companions approach each other. The outbursts of this type are weak (a maximum luminosity 
$L_{max}\simeq10^{37}$ erg s$^{-1}$) and short (lasting several days). The type II outbursts are irregular, longer (several weeks or months), and more powerful with $L_{max} \ge 10^{38}$ erg s$^{-1}$. The nature of the latter is not completely clear, but it is presumably related to the evolution of the decretion disk around the Be star \citep{negueruela01,okazaki01,moritani13,martin14,monageng17}. As a rule, the neutron star spins up during outbursts and spins down between them  \citep[see, e.g.,][]{postnov15}.

In this paper we investigate the dependence $\dot{\nu}(L_{bol})$ for eight X-ray pulsars in order to determine the exponent $\beta$ in the dependence $\dot{\nu}(\dot{M})$ and to obtain constraints on the magnetospheric size from the best observational data available to date. In   Section "Data Analysis" we describe the sample of pulsars that was used in this paper and the criteria that were applied to produce it. In  Section "Size of the Neutron Star Magnetosphere" we present the best fits to the dependence $\dot{\nu}(L_{bol})$ for several models of angular momentum transfer during the accretion of matter onto a neutron star that were used to determine the radius of the neutron star magnetosphere. In  Section "Estimating the Error
in the Magnetospheric Radius" we estimate the range of admissible magnetospheric radii related to the uncertainty in measuring the neutron star mass and radius and the distance to the system. The obtained results are discussed in  Section "Conclusions".

\section*{DATA ANALYSIS}

In this paper we selected the pulsars in Be systems satisfying the following criteria:
\begin{itemize}
\item  the demonstration of type II  outbursts, because such outbursts allow to measure the dependence $\dot{\nu}(L_{bol})$ in a wide range of luminosities and, consequently, mass accretion rates;
\item there are Fermi/GBM measurements \citep{meegan09} of the neutron star spin frequency during the outbursts;

\item  the orbital parameters of the system needed to correct the neutron star spin frequency for the orbital motion are known;

\item a cyclotron line is detected in the energy spectrum of the source, which allows to estimate the magnetic
field of the neutron star;
 
\item there are measurements of the distance to the system.
\end{itemize}

As a result, we obtained a list of eight objects, which is given in Table \ref{list_systems}. This table also provides the parameters of the binary systems and neutron stars used in our analysis and gives references to the papers
where these parameters were estimated. 

The neutron star magnetic field was determined from the cyclotron line energy using the formula
\[
B_{12}=E_{cycle}/11.6\times(1+z),
\]
where $B_{12}$ is expressed in $10^{12}$ Gauss, $E_{cycle}$ is the cyclotron energy in keV, and z is the gravitational
redshift. We used a simple approximation that the X-ray emission is generated near the surface of
a neutron star with standard parameters $M_{NS} = 1.4 \msun$ and $R_{NS} = 10^6$ cm; therefore, 
$z=(1-2GM_{NS}/c^2/R_{NS})^{-1/2}-1=0.3$. The stellar magnetic moment was calculated from the formula $\mu=BR^3/2$.

\begin{table*}
\caption{The list of Be systems investigated in this paper. The sources are arranged in order of increasing pulsation period }\label{list_systems}

\setlength{\extrarowheight}{5pt}

\begin{tabularx}{2.0\columnwidth}{l|l|C|C|C|C|C|C}
\hline
&Source& $d$, kpc& $E_{cycle}$, keV&B, $10^{12}$ Gauss &$P_{spin}$, c&$P_{orb}$, days&References\\

\hline
1&4U 0115+634&$8\pm1$&16	&$1.8\pm0.1$&3.61	&	24.32	& 1,2\\	
2&V 0332+53	&7 (6-9) &28.9	&$3.2\pm0.02$	&4.37	&	33.833	&3,4\\
3&RX J0520.5-6932	&$50\pm2$&31.	&$3.5\pm0.1$&8.04&	23.93	&5,6\\
4&2S 1553-542&$20\pm4$&23.5	&$2.9\pm0.2$&9.28	&31.345		&7,8\\		
5&XTE J1946+274	&$9\pm1$& 38.3	&$4.3\pm0.3$&15.75&	172.7	&9,10\\
6&KS 1947+300&$10.4\pm0.9$ & 12	&$1.37\pm0.08$&	18.81	&40.415&11,12\\
7&GRO J1008-57&$5.8\pm0.5$& 78 &	$8.7\pm0.3$	&93.5	&	249.48&11,13\\
8&1A0535+26	&2 (1.3-2.4)& 42.9	&	$4.8\pm0.1$&	103.5	&111.1	&14,15\\
\hline					
\end{tabularx}
\\

[1] \cite{negueruela01};  [2] \cite{tsygankov07}; [3] \cite{negueruela99}; [4] \cite{tsygankov06}; [5] \cite{inno13}; [6] \cite{tendulkar14}; [7] \cite{lutovinov16}; [8] \cite{tsygankov16}; [9] \cite{verrecchia02}; [10]\cite{doroshenko17b}; [11] \cite{Riquelme12}; [12] \cite{fuerst14}; [13] \cite{bellm14}; [14] \cite{steele98}; [15] \cite{maitra13}.
\end{table*}

\subsection*{Determining the Bolometric Luminosity}

To determine the bolometric luminosity of a source, we used the Swift/BAT data in the 15-50 keV energy band \citep{barthelmy05}. By applying the bolometric correction, we recalculated the flux in this energy band to the energy range 0.1-100 keV, which may be deemed bolometric for an accreting neutron star with a good accuracy. The BAT measurements
were recalculated to the luminosity via the known count rate for the Crab Nebula (0.22 counts\,cm$^{-2}$c$^{-1}$)
and the flux of 1.4$\times$10$^{-8} $ erg\, cm$^{-2}$c$^{-1}$,

\begin{equation} 
 L  = 4\pi d^{2} \left( \frac{1.4\times10^{-8}\, k_{bol}\,R_{BAT}}{0.22} \right),
 \end{equation}
where d is the distance to the source and $R_{BAT}$ is the BAT count rate.

\begin{figure*}
\begin{picture}(250,300)
\put(25,0){\includegraphics[width=1.8\columnwidth]{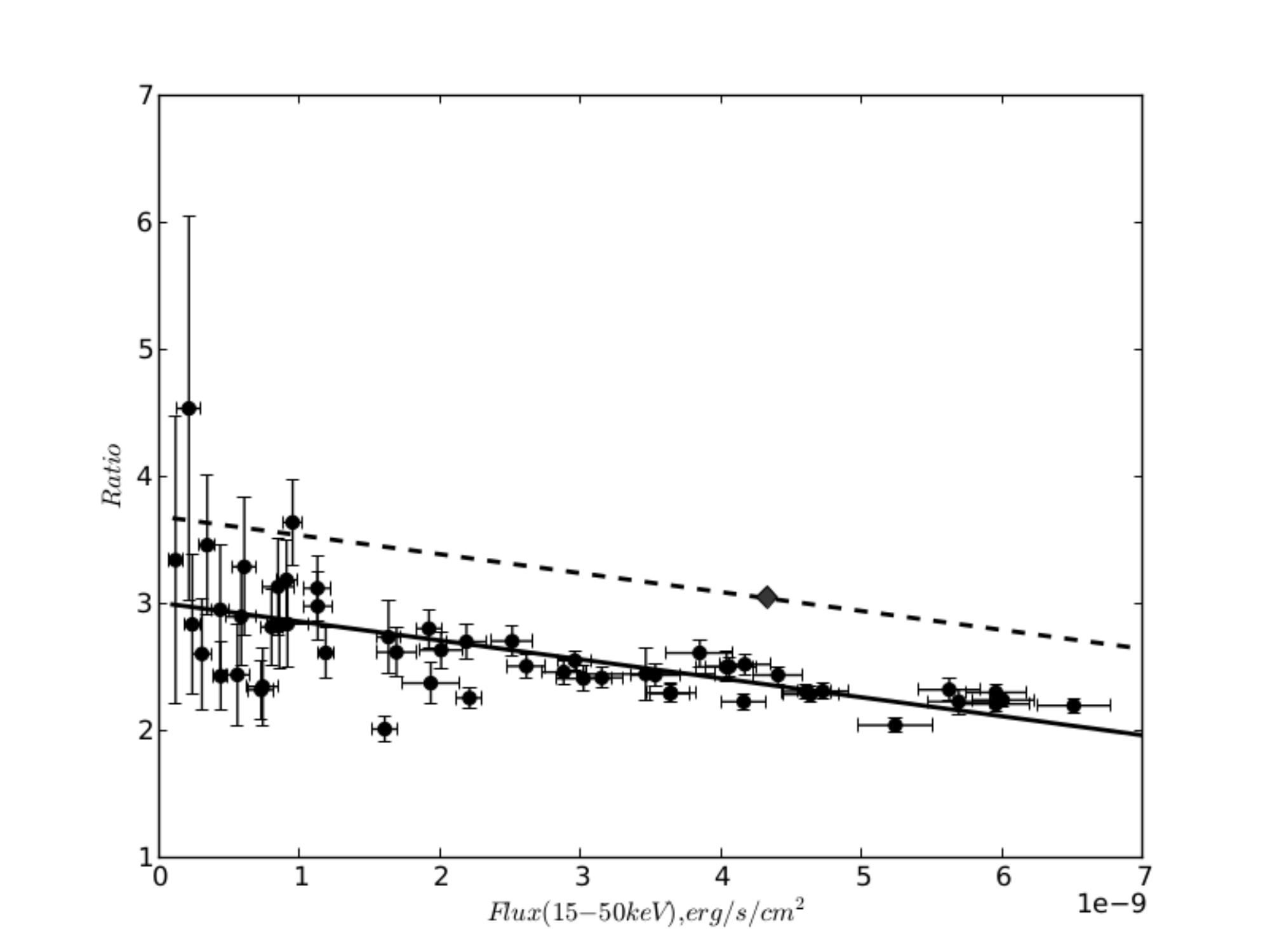}}
\put(320,260){\4u0115}
\end{picture}
\caption{The filled circles indicate the dependence of the flux ratio $(F_{2-10}+F_{15-50})/F_{15-50}$ ($(F_{0.3-10}+F_{15-50})/F_{15-50}$ for \ks1947) on the 15-50 keV flux during the outbursts of the systems being investigated in this paper. The solid line represents the best-fit model. The diamonds indicate the flux ratio $(F_{0.1-100}/F_{15-50})$  derived from a spectral model or the analysis of broadband spectra. The dashed line indicates the flux dependence of the bolometric correction, which are the above best-fit models scaled to the measured flux ratios $F_{0.1-100}/F_{15-50}$. These dependences were used in the subsequent analysis.
}\label{bol_kor}
\end{figure*}

The bolometric correction $k_{bol}$ and its luminosity dependence were calculated for each source separately by the method proposed by  \cite{tsygankov17}. For this purpose, apart from the BAT data, we used the MAXI data in the 2-10 keV energy
band \citep{matsuoka09} and the Swift/XRT data in the 0.3-10 keV energy band processed by means of the Online Service \citep{evans09} for \ks1947. To determine the dependence of the bolometric correction on the 15-50 keV flux, we constructed the following dependences of the flux ratio in different energy bands: $(F_{2-10}+F_{15-50})/F_{15-50}$ for the MAXI data and $(F_{0.3-10}+F_{15-50})/F_{15-50}$ for the Swift/XRT data (Fig. \ref{bol_kor}). When constructing these dependences, we used the Swift/BAT data in which the source detection significance exceeded $2\sigma$. To determine the normalization of the dependence of $k_{bol}=F_{0.1-100}/F_{15-50}$ on the 15-50 keV flux, we used either the data on the broadband spectrum during the source outburst activity from the literature or the broadband spectra constructed from
the Swift/XRT and Swift/BAT data directly for the outbursts being investigated (for \a0535 and \v0332). The derived dependences of the bolometric correction on the 15-50 keV flux are indicated in Fig. \ref{bol_kor} by the dashed line. These dependences were used in the subsequent analysis. For the system \2s1553 we failed to perform such an analysis,
because the Swift/BAT and MAXI source detection significance during the outburst was insufficient, $(2-5)\sigma$. The bolometric correction for this source was calculated from the data on the broadband spectrum from \cite{tsygankov16}. The derived bolometric corrections (Table \ref{bol_cor}) lie within the range 2-3.5, in good agreement with the shape of the spectra for X-ray pulsars in a wide energy range \citep{filippova05}.

\begin{figure*}
\begin{picture}(250,250)
\put(25,0){\includegraphics[width=1.8\columnwidth]{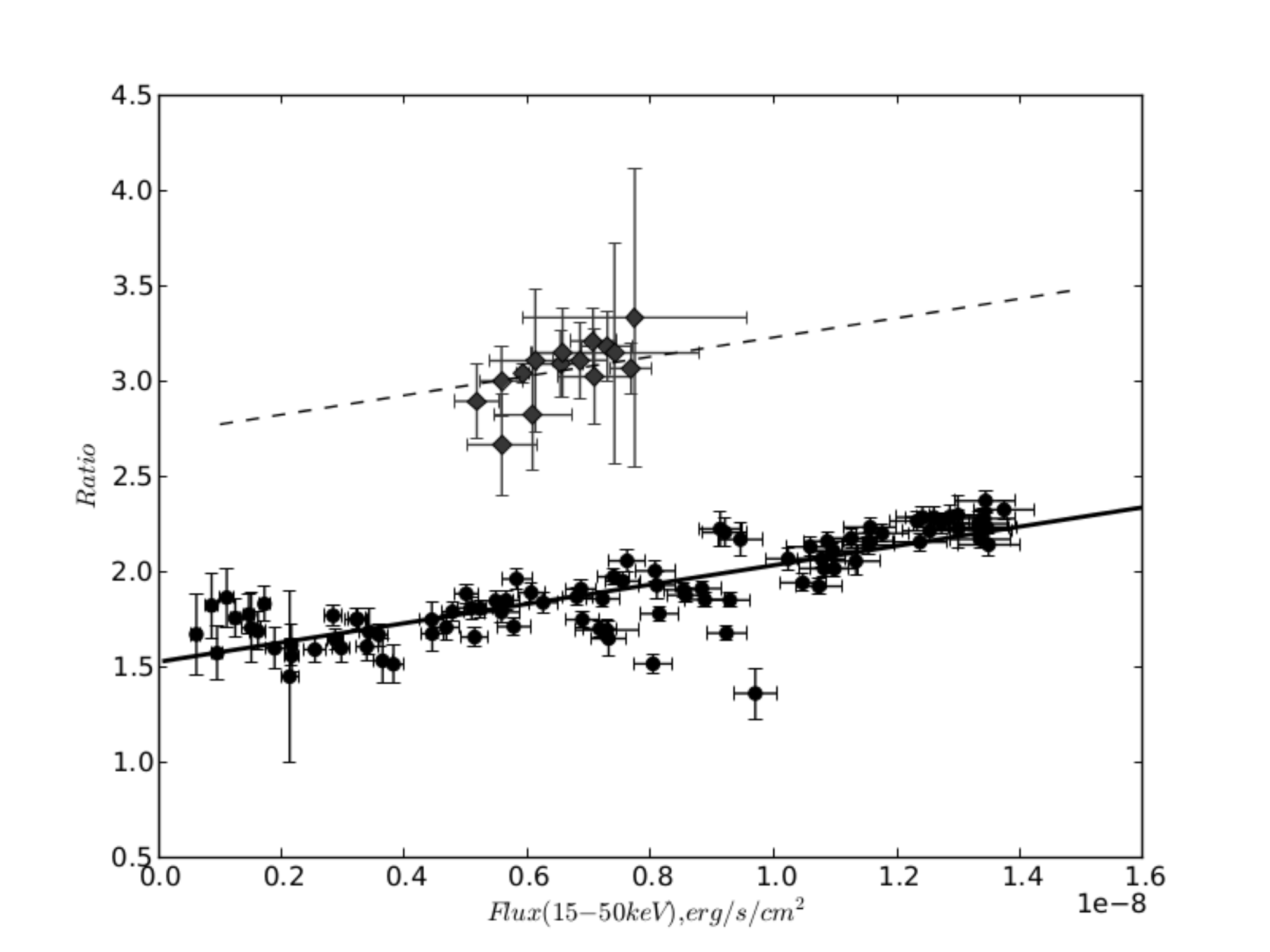}}
\put(320,260){\v0332}
\end{picture}
\\
\\
\\
\\
\\

\begin{picture}(250,250)
\put(25,0){\includegraphics[width=1.8\columnwidth]{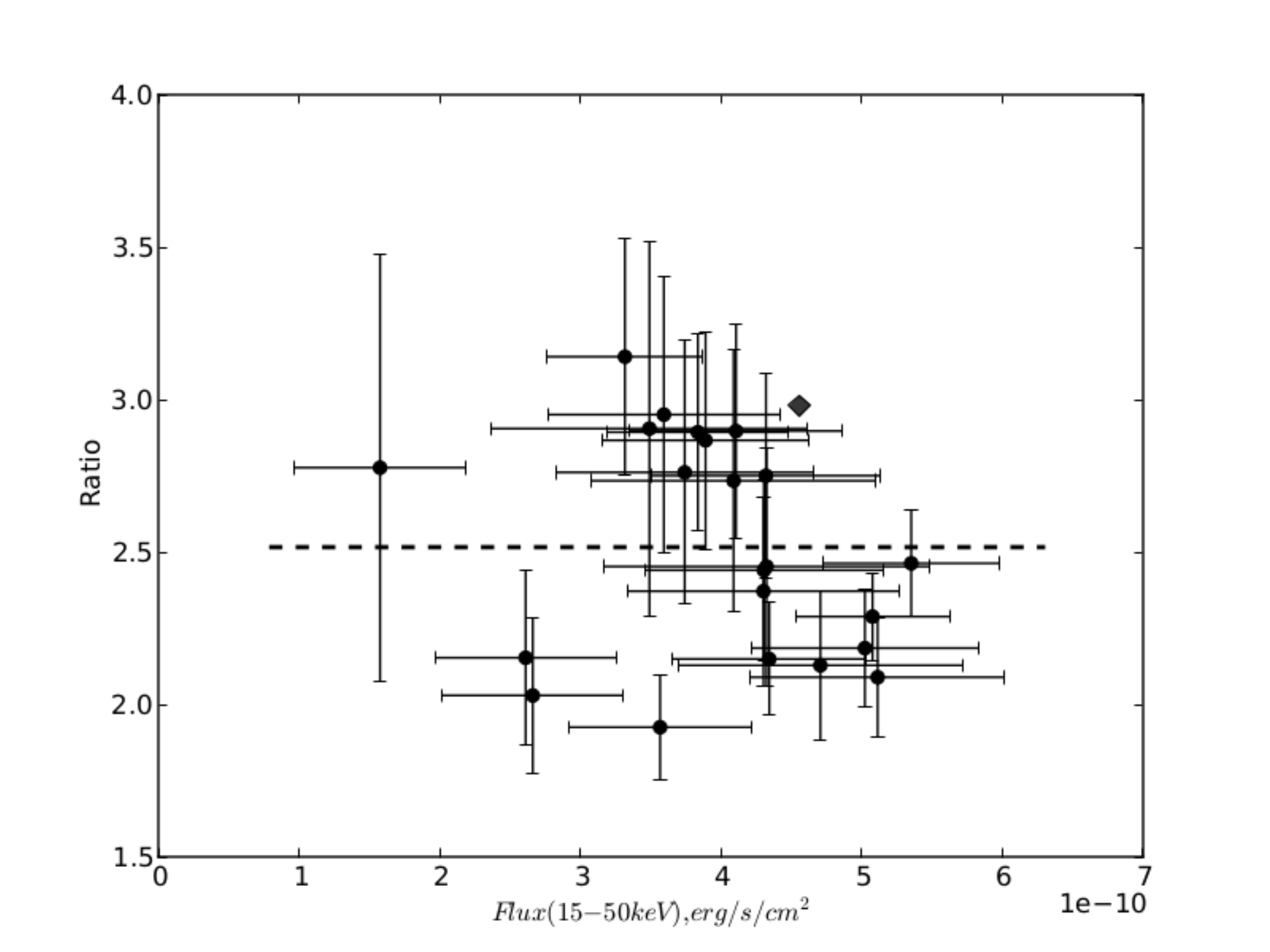}}
\put(320,260){\rx0520}
\end{picture}
\center{\small\textbf{Fig.~1} --- Contd.}
\end{figure*}

\begin{figure*}
\begin{picture}(250,250)
\put(25,0){\includegraphics[width=1.8\columnwidth]{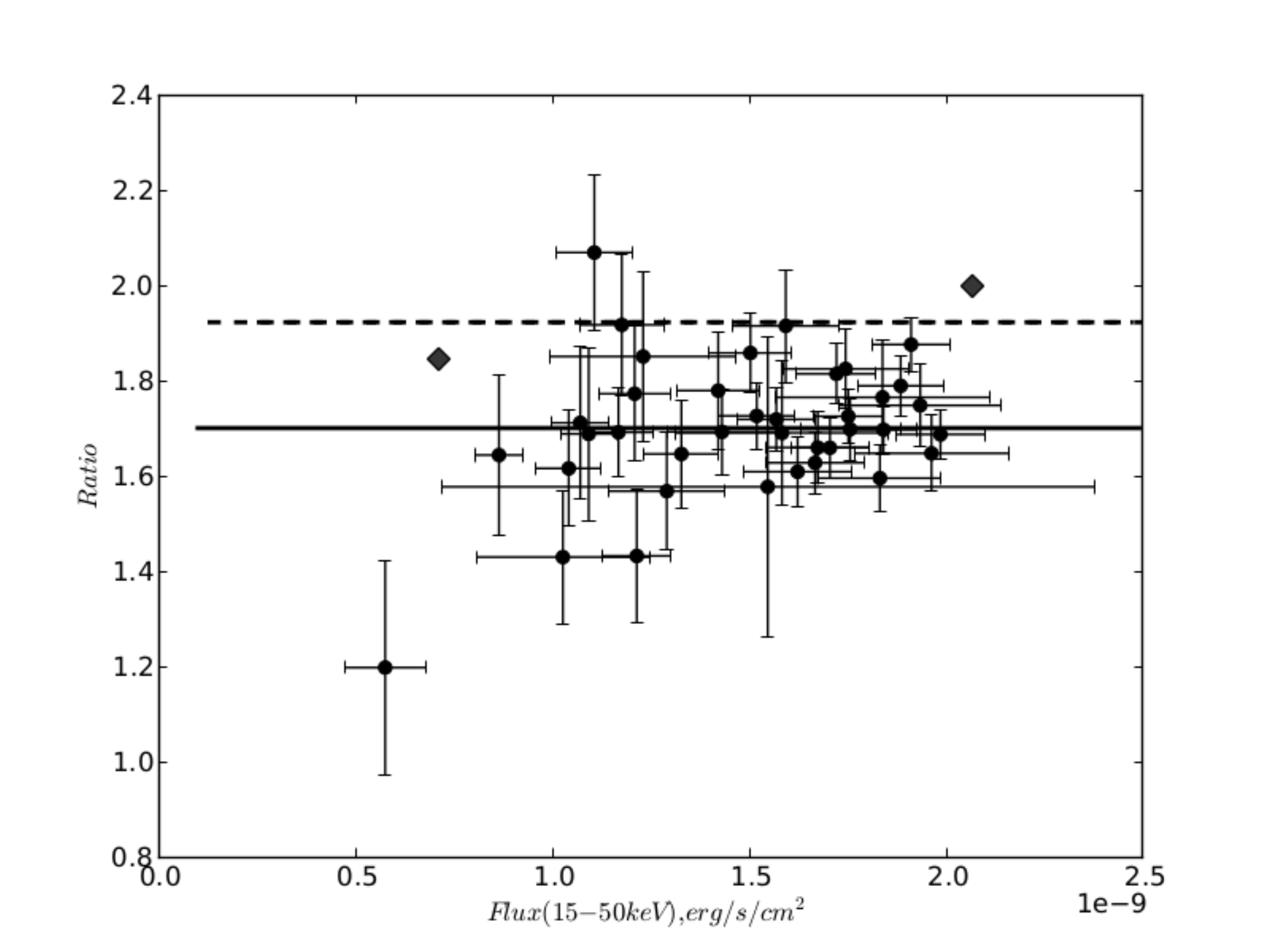}}
\put(320,260){\xte1946}
\end{picture}
\\
\\
\\
\\
\\

\begin{picture}(250,250)
\put(25,0){\includegraphics[width=1.8\columnwidth]{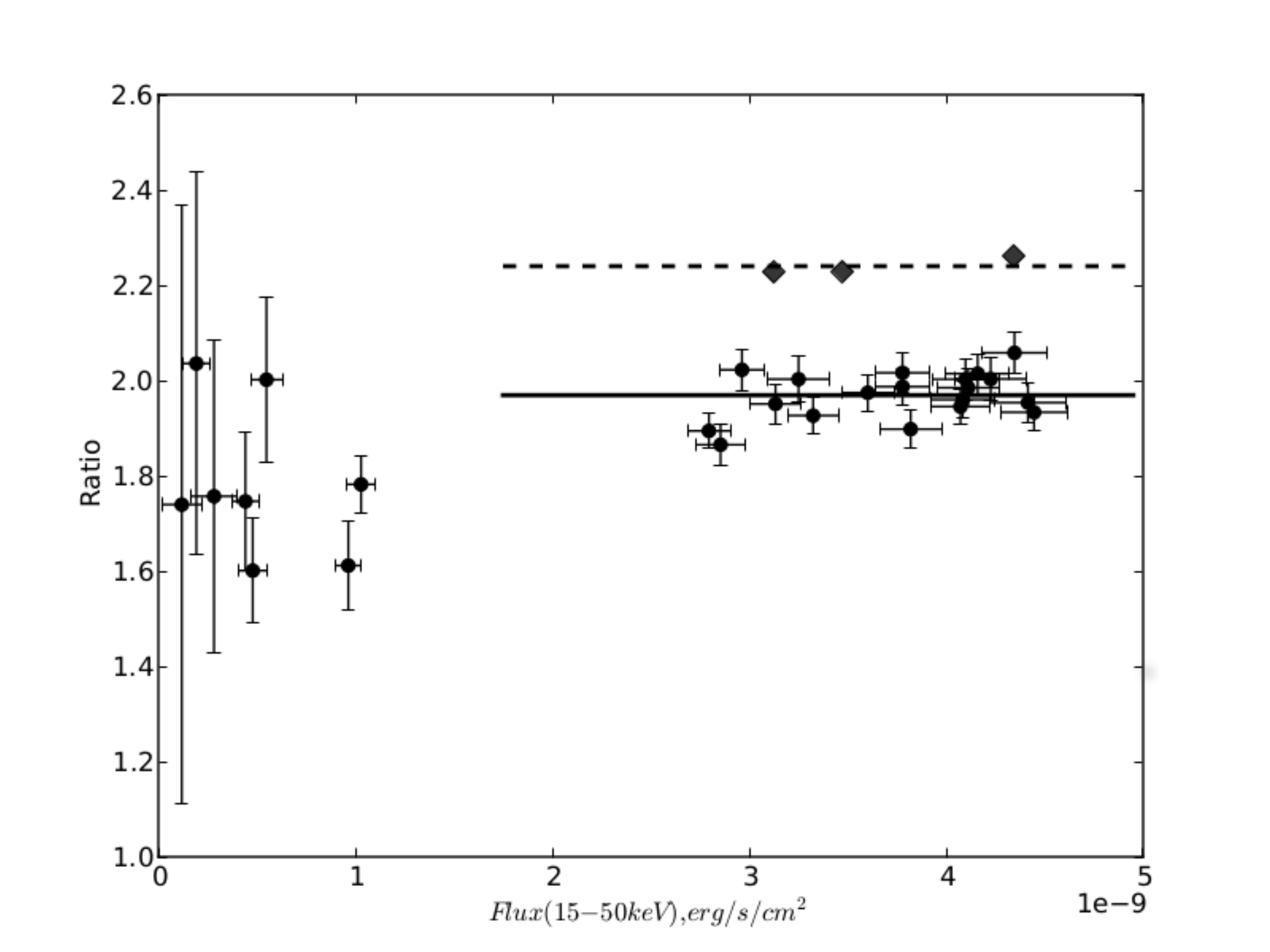}}
\put(320,260){\ks1947}
\end{picture}
\center{\small\textbf{Fig.~1} --- Contd.}

\end{figure*}

\begin{figure*}

\begin{picture}(250,250)
\put(25,0){\includegraphics[width=1.8\columnwidth]{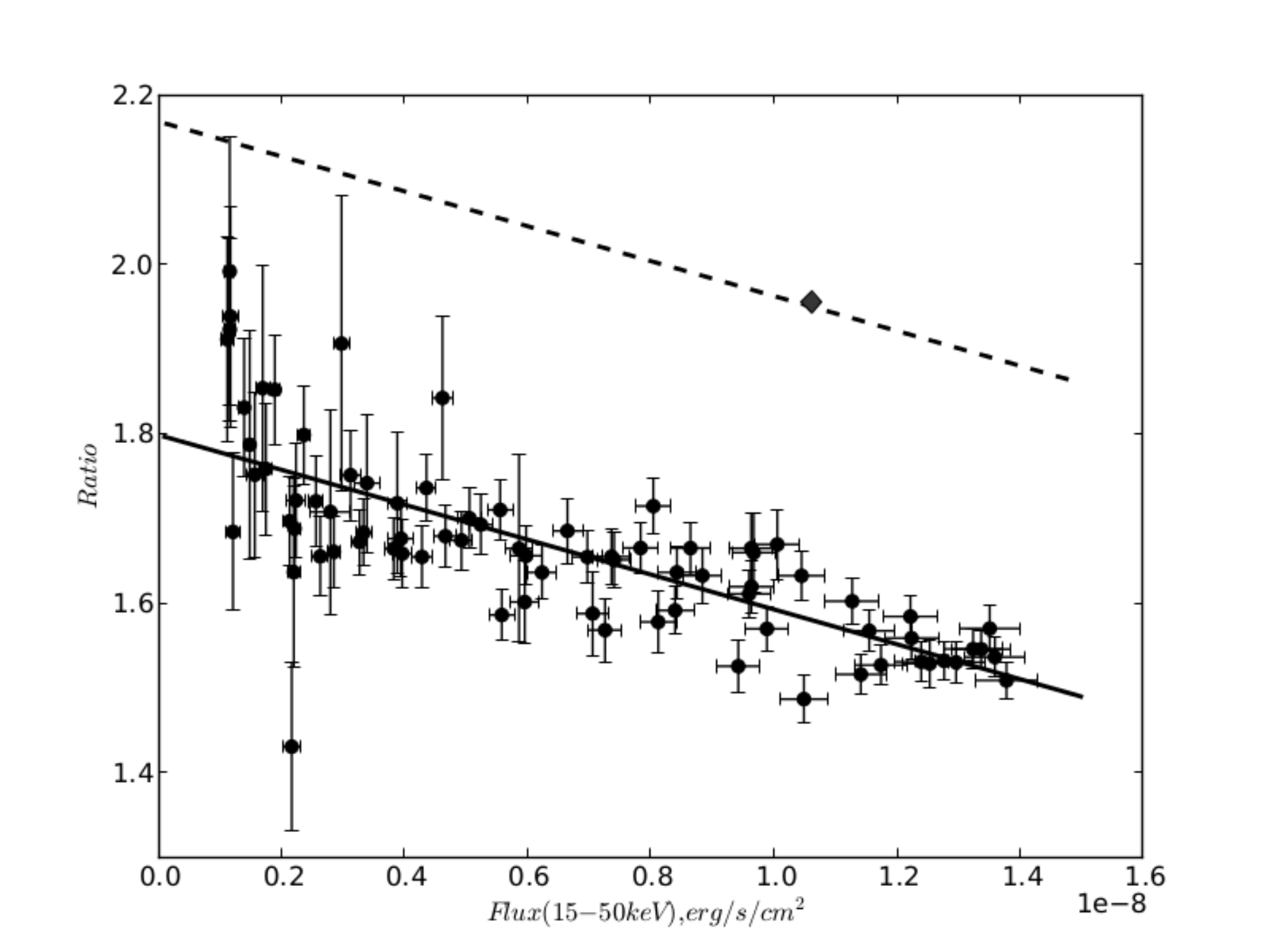}}
\put(320,260){\g1008}
\end{picture}
\\
\\
\\
\\
\\

\begin{picture}(250,250)
\put(25,0){\includegraphics[width=1.8\columnwidth]{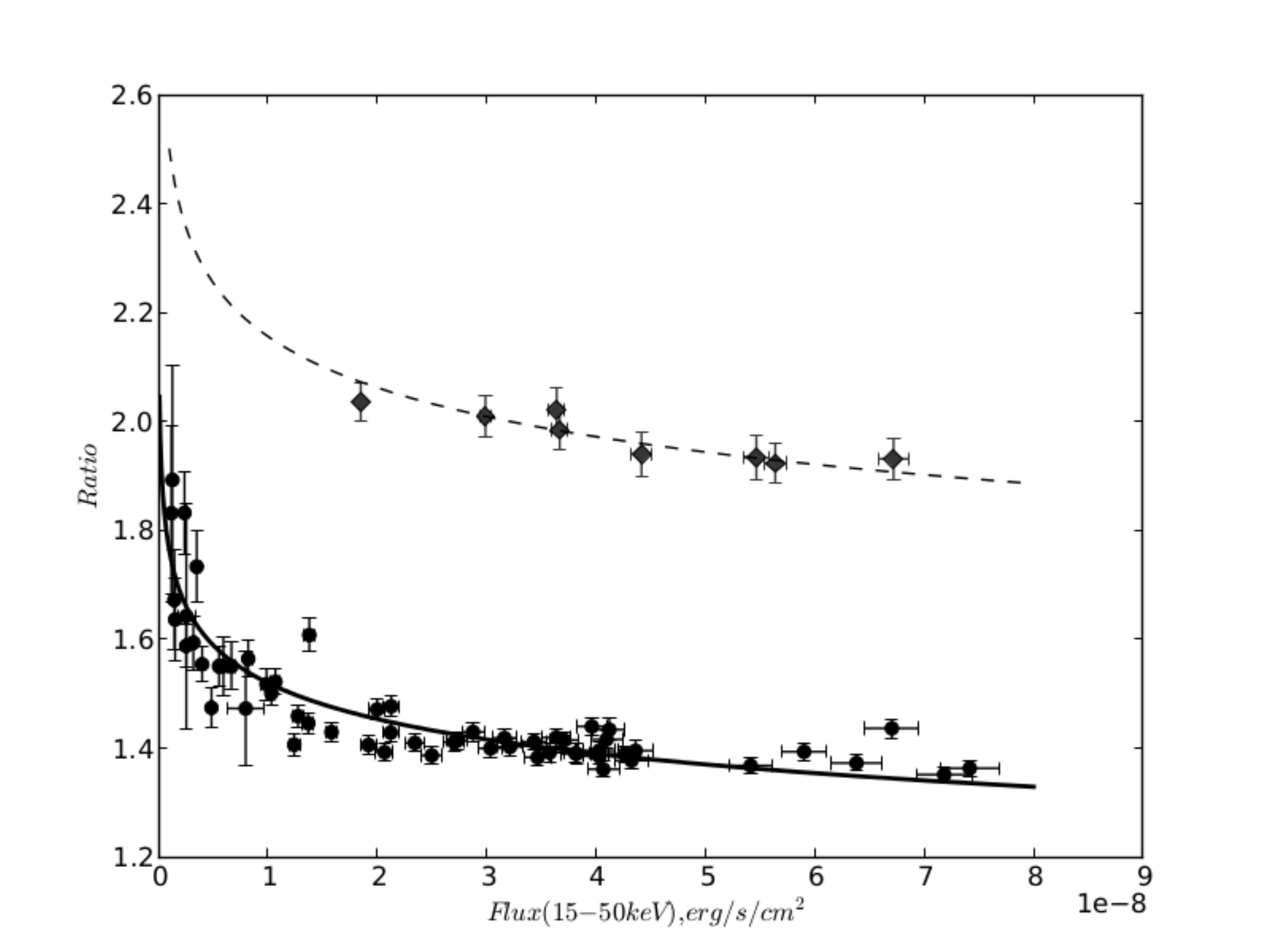}}
\put(320,260){\a0535}
\end{picture}

\center{\small\textbf{Fig.~1} --- Contd.}
\end{figure*}

\subsection*{Determining the Timing Characteristics}

For two systems, \g1008 and \xte1946, the publicly available GBM data were not corrected for the orbital motion of the binary system. Therefore, we did this by ourselves using a standard technique \citep{argyle04} and the following
orbital parameters: the eccentricity $e=0.68$, the projected orbital semimajor axis $a\sin{i}=530$ lt-s,
the longitude of periastron $\omega=-26\degree$, the orbital period  P$_{orb}$=249.48 days, and the time of periastron passage $\tau$=54424.71 MJD for \g1008 \citep{coe07,kuhnel13} ; $e=0.246$, $a\sin{i}=471.2$ lt-s, $\omega=-87.4\degree$, P$_{orb}$=172.7 days and $\tau$=55514.8 MJD for \xte1946 \citep{marcu15}.

\begin{figure*}
\begin{picture}(270,270)
\put(100,0){\includegraphics[width=1.8\columnwidth,bb=50 50 800 800]{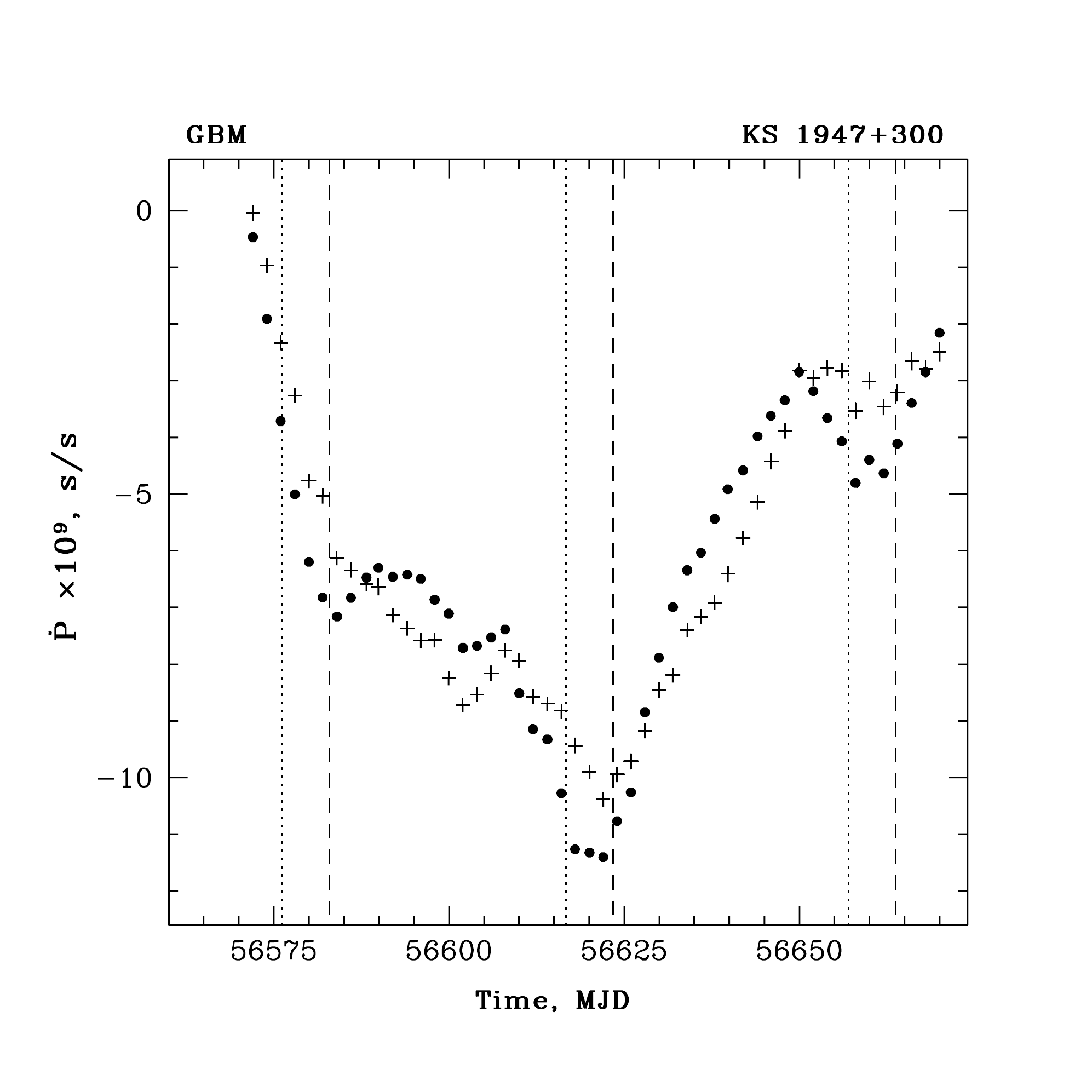}}
\end{picture}
\caption{Time dependence of the period derivative $\dot{P}$ for \ks1947 during the outburst being investigated here. The filled
circles indicate the dependence that was constructed from the data corrected for the orbital motion using the ephemeris from
\cite{galloway04}. The crosses indicate the dependence constructed using the orbital parameters improved in our paper.
The vertical lines indicate the times of periastron passage: in accordance with the orbital parameters from \cite{galloway04} (dotted lines) and according to the parameters obtained in this paper (dashed lines).
 }\label{ks1947_cor}
\end{figure*}

For the system \ks1947 the time dependence of the period derivative $\dot{P}$ constructed from the GBM\footnote{\url{https://gammaray.nsstc.nasa.gov/gbm/science/pulsars}} (the periods were corrected for the orbital
motion using the orbital parameters from \citep{galloway04}) exhibits features with a characteristic period equal to the orbital one (Fig. \ref{ks1947_cor}). Such a behavior may point to a deviation of the true orbital parameters at the time of our observations from those used to correct the neutron star spin frequency, i.e., to the fact that the orbital parameters could slightly
change since their determination in 2004. Using the data uncorrected for the orbital motion, we improved
the orbital ephemeris of the binary system at the time of our observations. The pulsation period could change as due to the Doppler shift  caused by orbital motion  so due to the accretion of angular momentum during the outburst.
 As has been noted in the Introduction, the spin frequency derivative is proportional to the accretion rate at some power at first approximation 6/7,  $\dot{P} \sim R_{BAT}^{6/7}$. That is a form that we used to find a solution for $\dot{P}$ measured by GBM and corrected for the orbital motion for a set of different orbital system parameters. In reality, it is difficult to improve all parameters of the binary system based on the data for only one outburst. Therefore, we varied  only one of them, the time of periastron passage $\tau$, to which the correction procedure is most sensitive. As a result, we found that the best solution corresponds to
$\tau = 56582.941$ MJD, which differs from the predictions made in accordance with the orbital parameters
from \citep{galloway04} by several days (Fig. \ref{ks1947_cor}). It can be seen from the figure that the dips in the time dependence of $\dot{P}$ from the original data correspond to the improved times of periastron passage, where maximum deviations from the true neutron star spin period derivatives are expected in the case of using incorrect orbital parameters.

As it is mentioned above both the Doppler shift and the intrinsic pulsar frequency change related to the accretion of matter should be taken into account during correction the frequency for the orbital motion. To a first approximation, the pulsation frequency change due to accretion
may be deemed proportional to the flux, $\dot{\nu}\sim F^{\beta}$. On the one hand, such an approach puts information into
the dependences $\dot{\nu}(L_{bol})$ constructed in this paper, but, on the other hand, the quality of the data (their
scatter and the error in $\dot{\nu}$) makes the dependence insensitive to this law. We checked this assertion using \xte1946 as an example, for which the orbital parameters are provided in the literature for two cases: $\beta$ = 6/7 and 1 \citep{marcu15}. The
difference in $\dot{\nu}$ for different $\beta$ turned out to be considerably smaller than the measurement errors, while fitting the dependence by a power law gave the same exponent for both cases.

The GBM data are rarer than the BAT data, i.e., one pulsation period measurement corresponds to several flux measurements. The longest time step between GBM observations we used is 3.5 days. To obtain the sets of interested parameters (t;L;$\dot{\nu}$), we took the measurements of the frequency $\nu$ (corrected for the orbital motion) from the Fermi/GBM data. For each pair of successive ${\nu}_i$, ${\nu}_{i+1}$ measurements made at times $t_i,t_{i+1}$ and spaced less than 3.5 days apart we selected all of the daily
averaged Swift/BAT observations that fell into this time interval; there can be $j = 0, 1, 2, 3$ such points. If only one observation fell into the above interval, then $\dot{\nu}=(\nu_{i+1}-\nu_i)/(t_{i+1}-t_i)$ was assigned to the observed luminosity $L$. If there were two points (with recording times $tb_j,tb_{j+1}$) and if both points lay on one side of $t_{mean}=(t_{j+1}+t_j)/2$, then we wrote the set $((tb_{j+1}+tb_j)/2;L_w;\dot{\nu})$, where $L_w$ is the weighted mean. If, alternatively, $tb_j,tb_{j+1}$ lay on different sides of $t_{mean}$ or if there were three points, then, in this case, a straight line was drawn through them by the least squares method (given the errors in the luminosity) and then the interpolated value $L_{interp}$ was determined at $t_{mean}$. 

\begin{figure*}
\begin{picture}(250,250)
\put(25,0){\includegraphics[width=2\columnwidth]{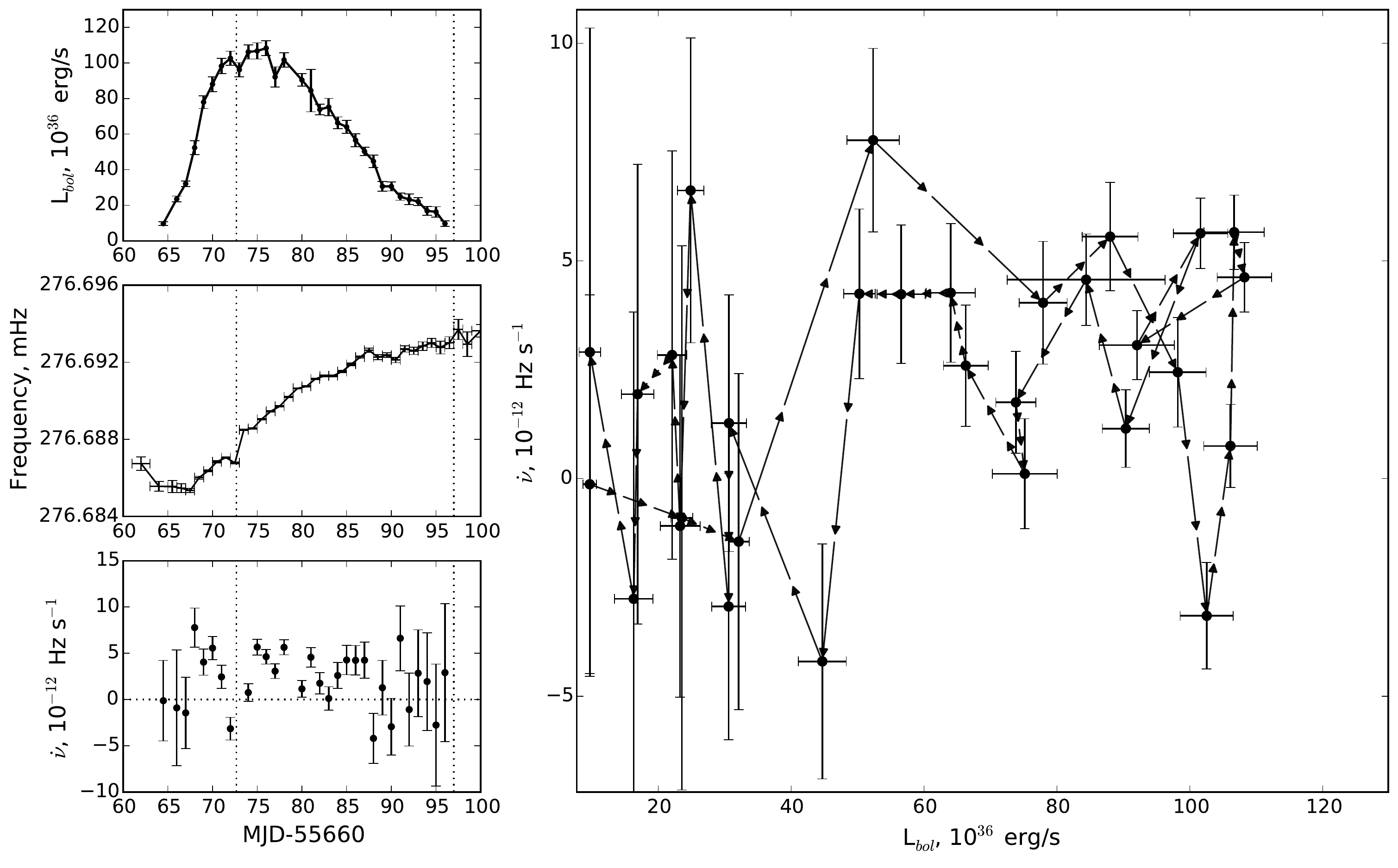}}
\put(400,270){\4u0115}
\end{picture}
\\
\\
\\
\\
\\

\begin{picture}(250,250)
\put(25,0){\includegraphics[width=2\columnwidth]{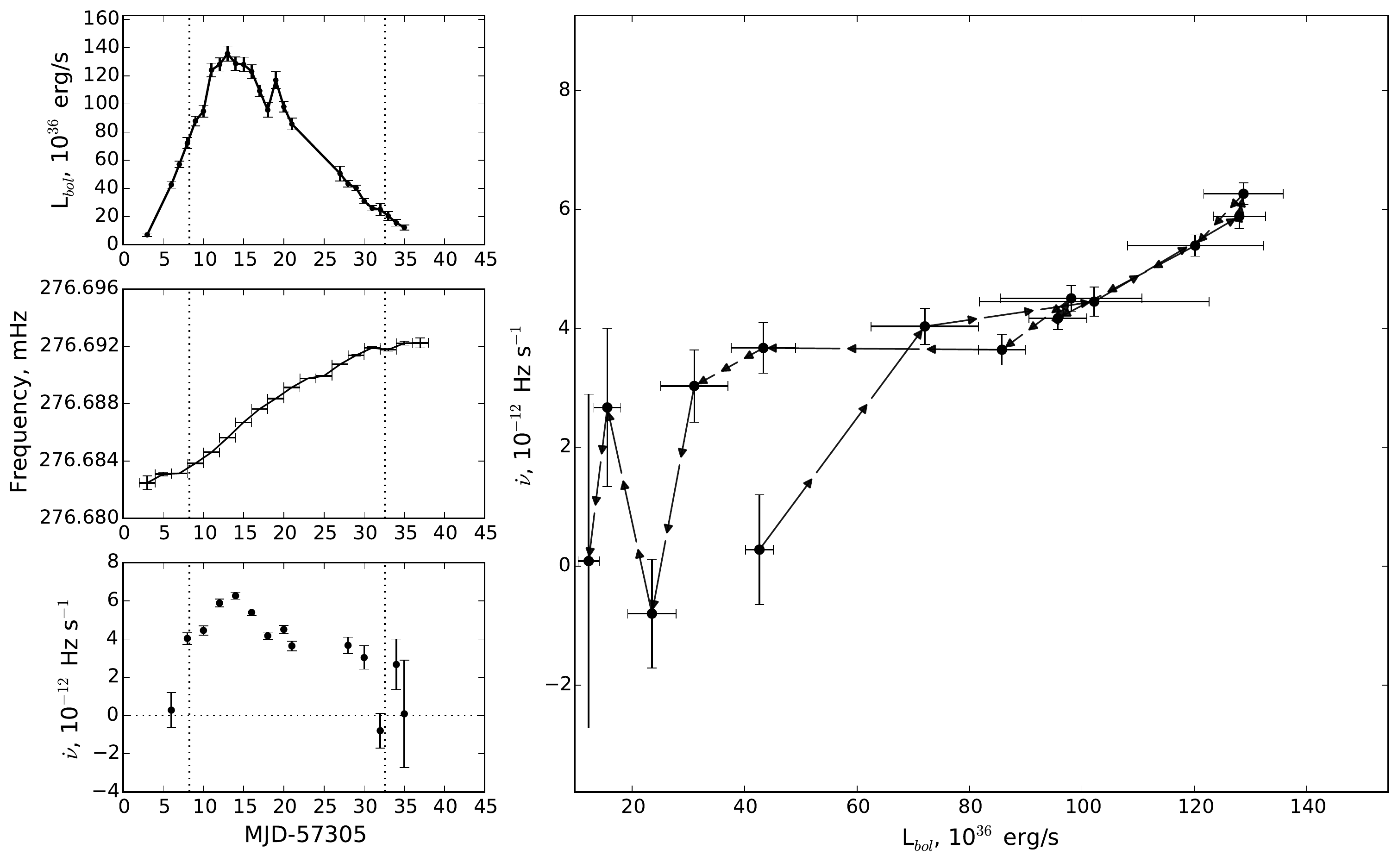}}
\put(400,270){\4u0115}
\end{picture}
\caption{Evolution of the parameters of the systems during their outbursts. For each system the figure shows on the left, from
top to bottom: the light curve, the time dependence of the spin frequency and the spin frequency derivative (the vertical dashed
lines mark the times of periastron passage); the observed dependence $\dot{\nu}(L_{bol})$ is shown on the right. The arrows indicate the
direction in which the system passes from point to point with time.}\label{observ}
\end{figure*}

The errors in the luminosity were calculated as a sum of the error in the weighted mean and the half of luminosity measurement range. This provided physically motivated results both in the regions of low fluxes, where the error in the weighted mean dominates, and at the outburst maxima, where averaging several points with a high significance leads to an underestimation of the actual luminosity range in
which $\dot{\nu}$ was measured. Figure \ref{observ} shows the time dependences of the luminosity, spin frequency, and spin frequency derivative for the outbursts investigated in this paper.

\begin{figure*}
\begin{picture}(250,250)
\put(25,0){\includegraphics[width=2\columnwidth]{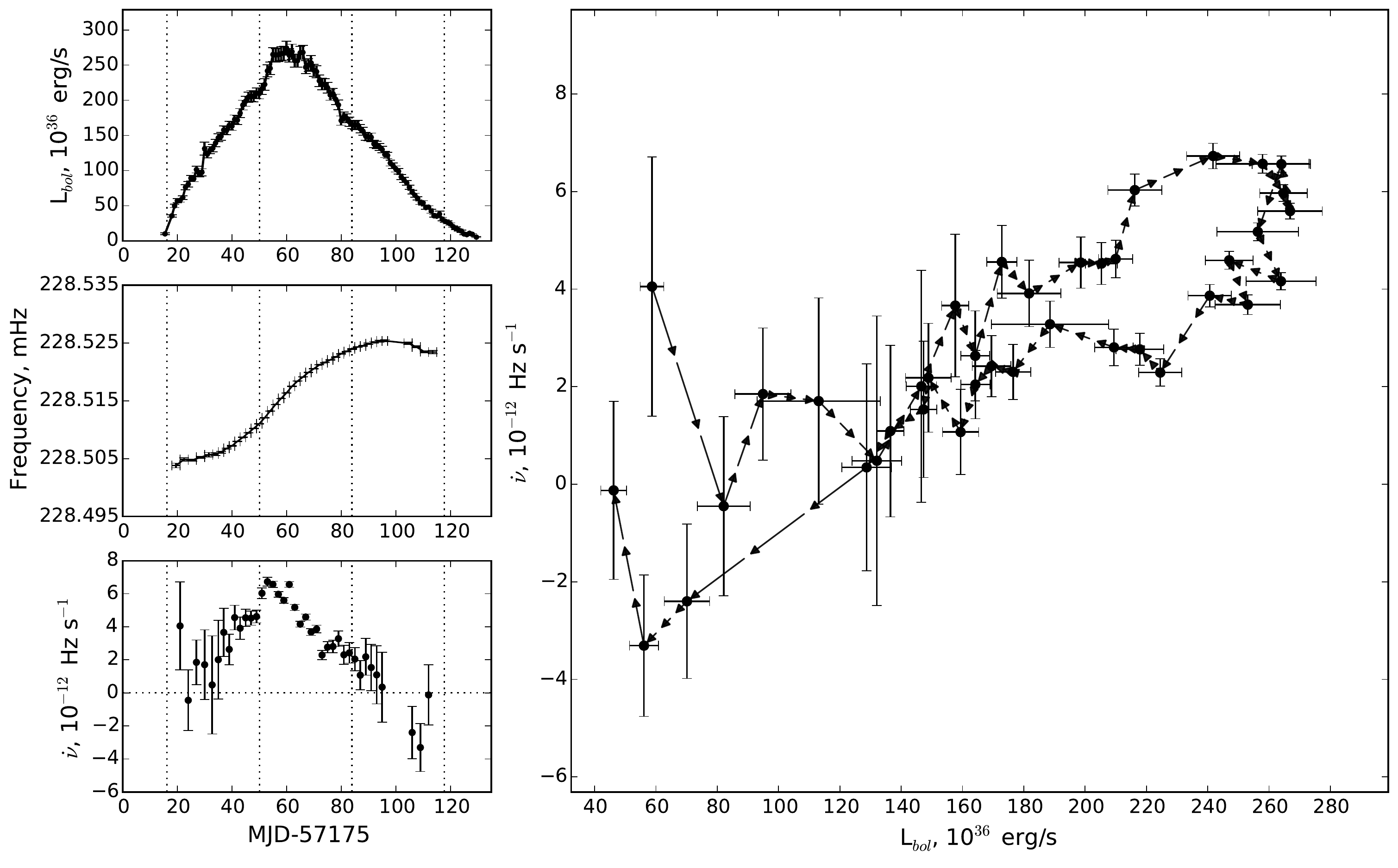}}
\put(400,270){\v0332}
\end{picture}
\\
\\
\\
\\
\\

\begin{picture}(250,250)
\put(25,0){\includegraphics[width=2\columnwidth]{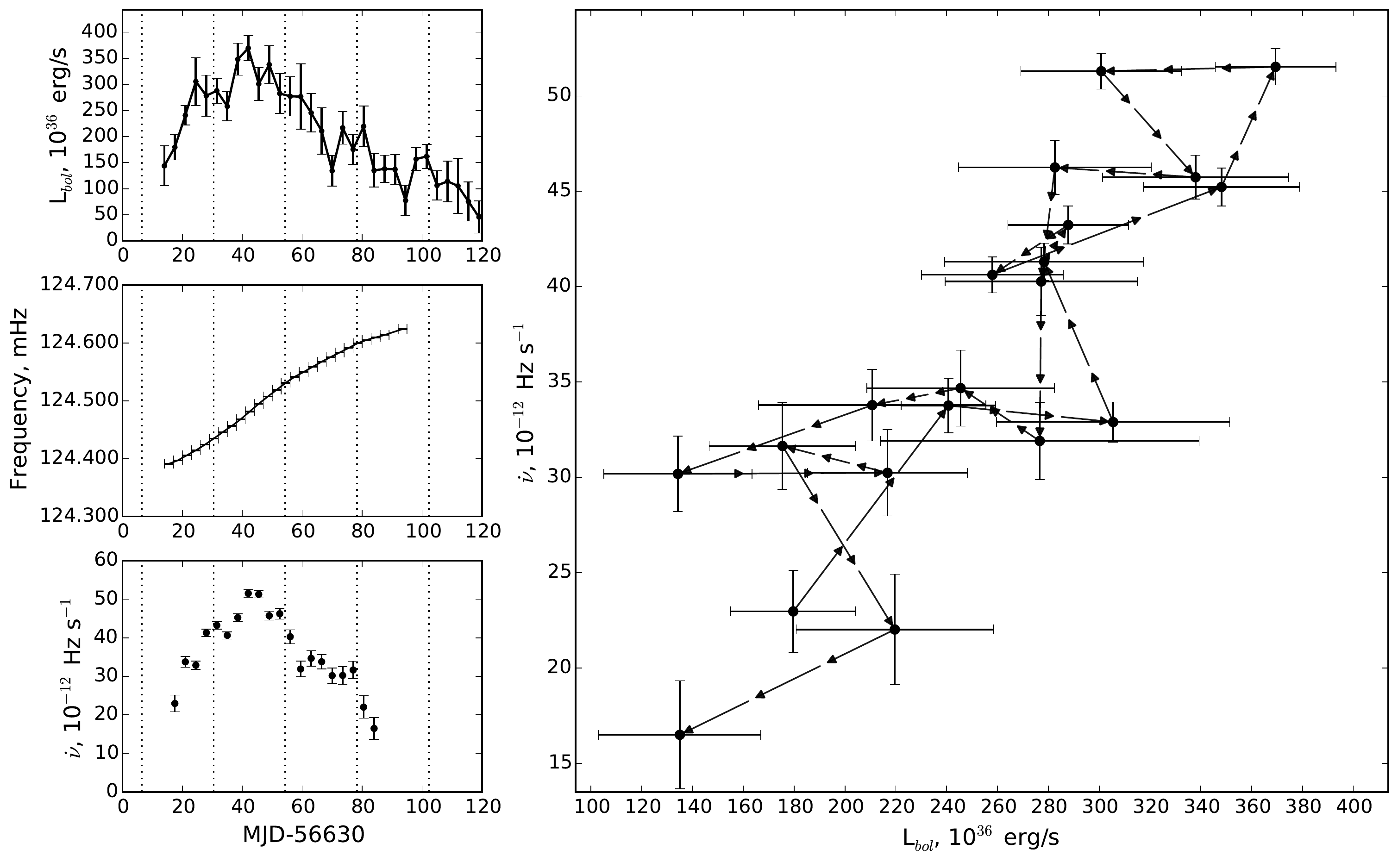}}
\put(250,270){\rx0520}
\end{picture}
\center{\small\textbf{Fig.~\ref{observ}} --- Contd.}
\end{figure*}

\begin{figure*}
\begin{picture}(250,250)
\put(25,0){\includegraphics[width=2\columnwidth]{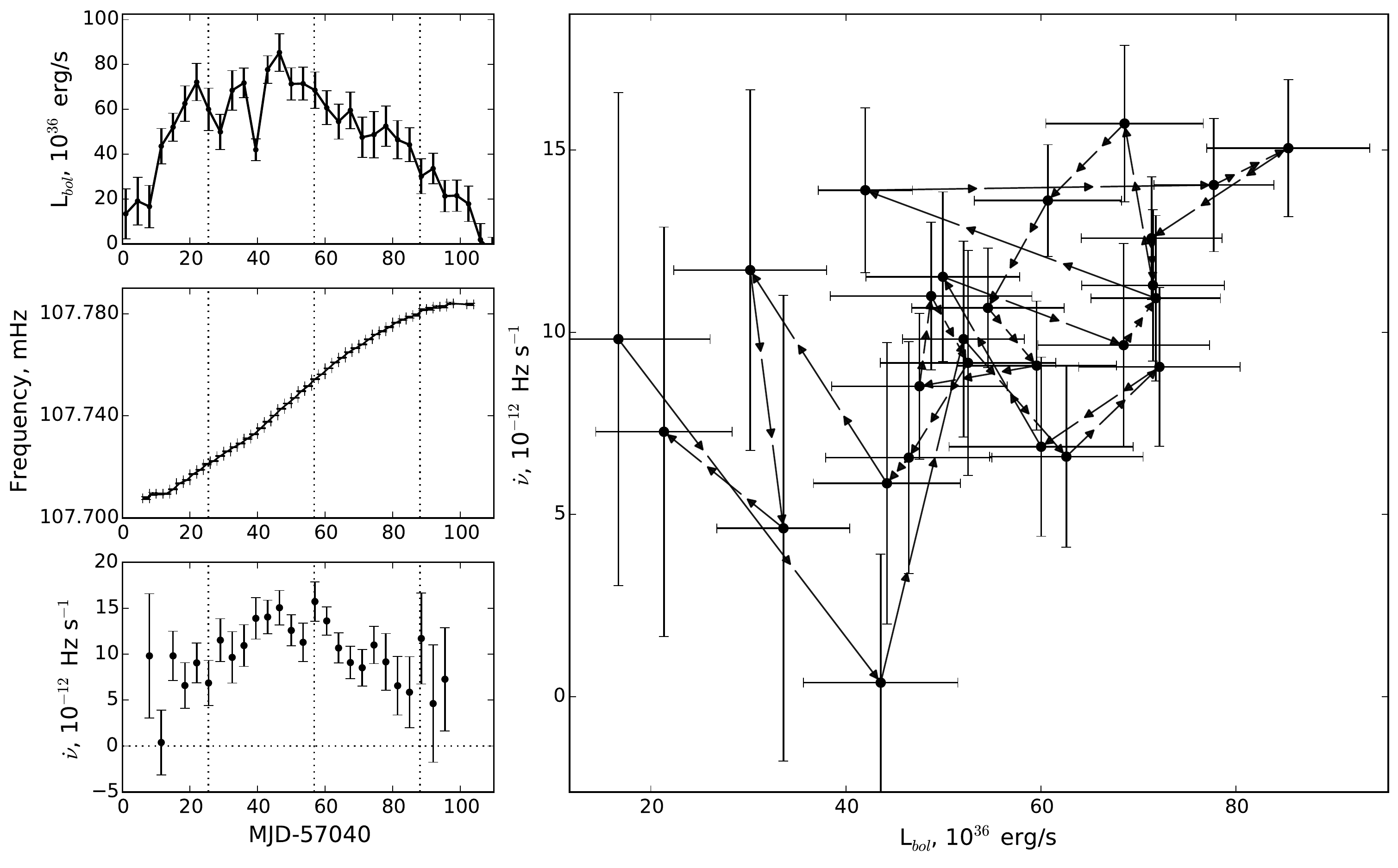}}
\put(250,270){\2s1553}
\end{picture}
\\
\\
\\
\\
\\

\begin{picture}(250,250)
\put(25,0){\includegraphics[width=2\columnwidth]{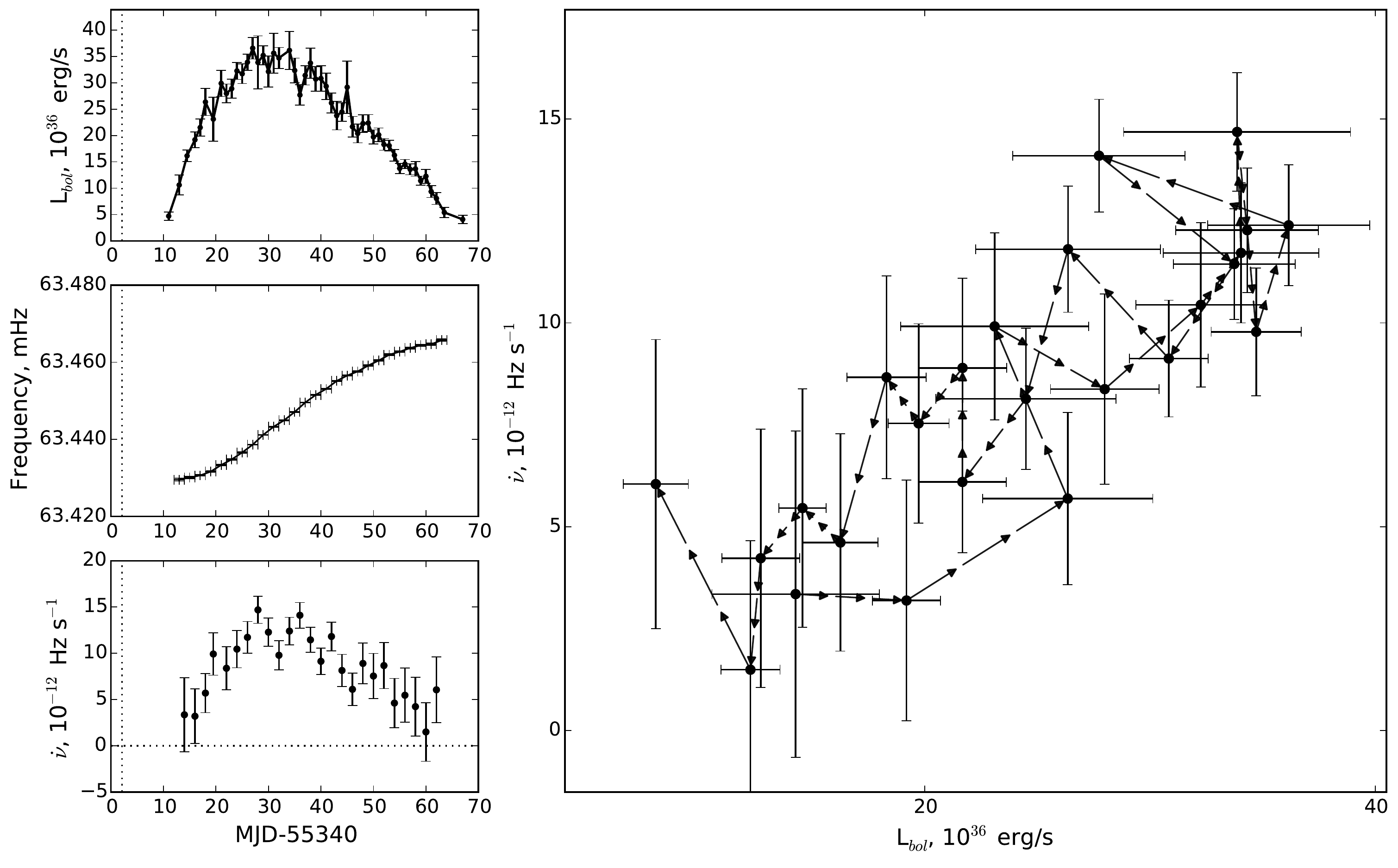}}
\put(250,270){\xte1946}
\end{picture}

\center{\small\textbf{Fig.~\ref{observ}} --- Contd.}
\end{figure*}

\begin{figure*}
\begin{picture}(250,250)
\put(25,0){\includegraphics[width=2\columnwidth]{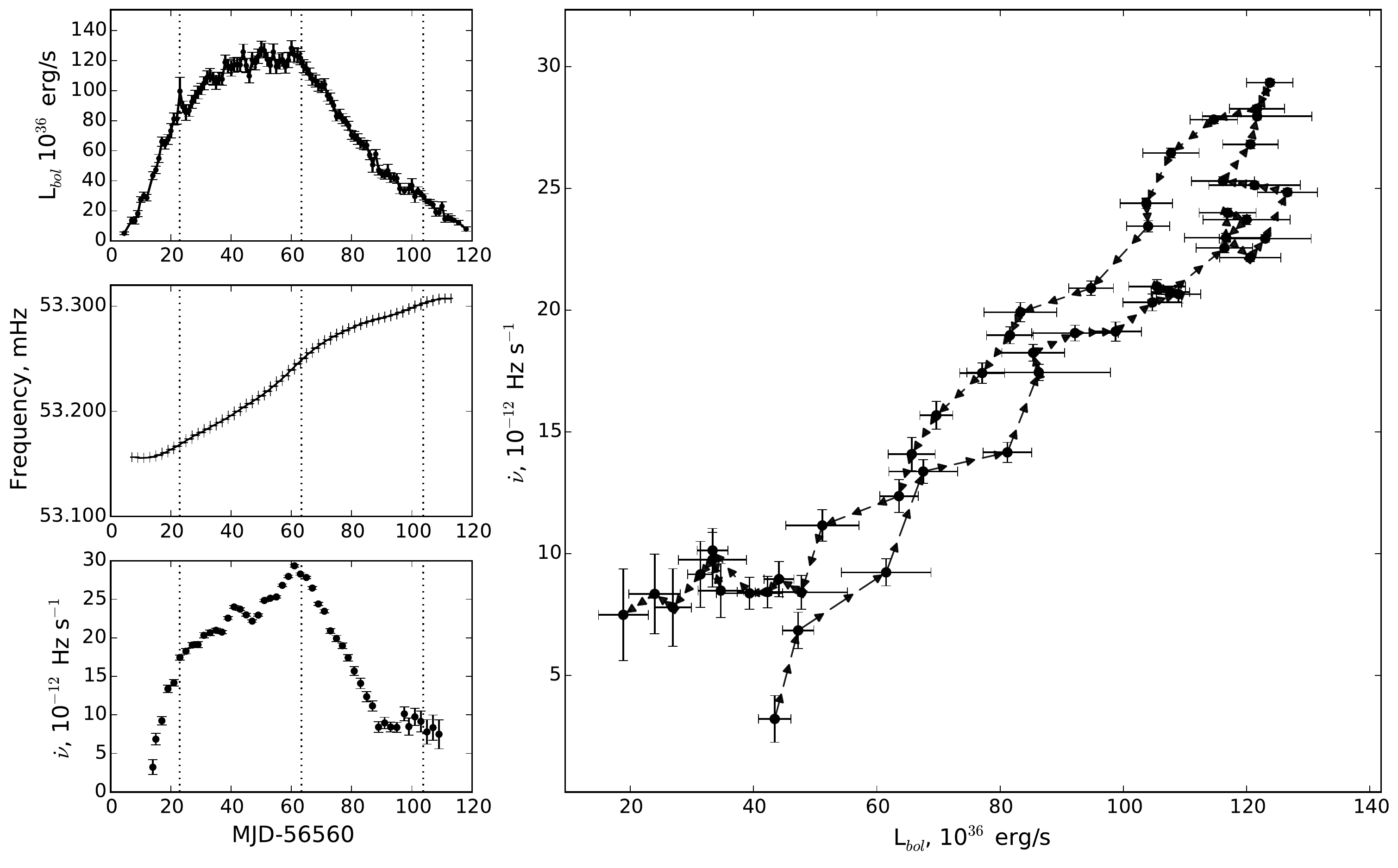}}
\put(250,270){\ks1947}
\end{picture}
\\
\\
\\
\\
\\

\begin{picture}(250,250)
\put(25,0){\includegraphics[width=2\columnwidth]{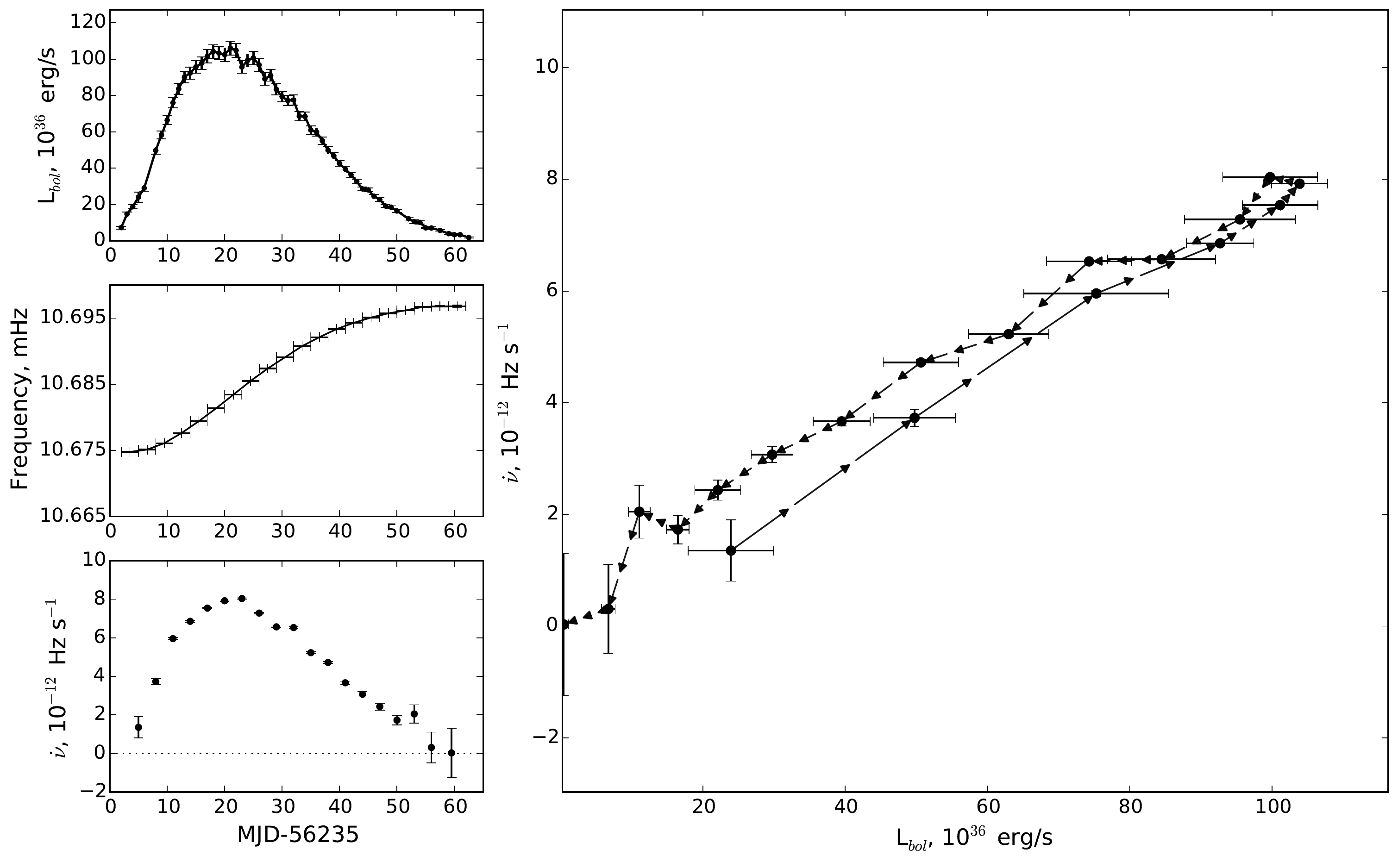}}
\put(250,270){\g1008}
\end{picture}
\center{\small\textbf{Fig.~\ref{observ}} --- Contd.}
\end{figure*}

\begin{figure*}
\begin{picture}(250,250)
\put(25,0){\includegraphics[width=2\columnwidth]{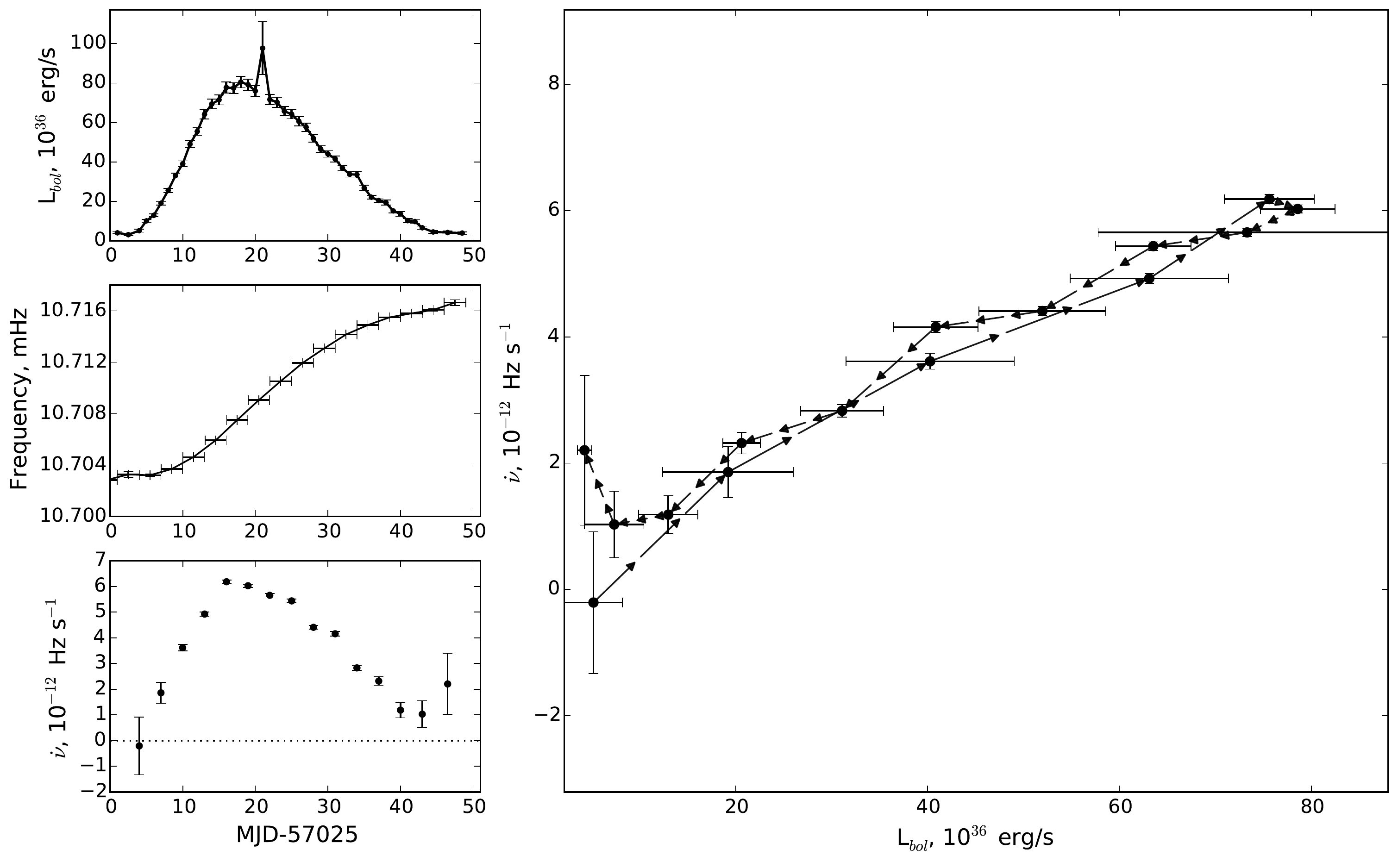}}
\put(250,270){\g1008}
\end{picture}
\\
\\
\\
\\
\\

\begin{picture}(250,250)
\put(25,0){\includegraphics[width=2\columnwidth]{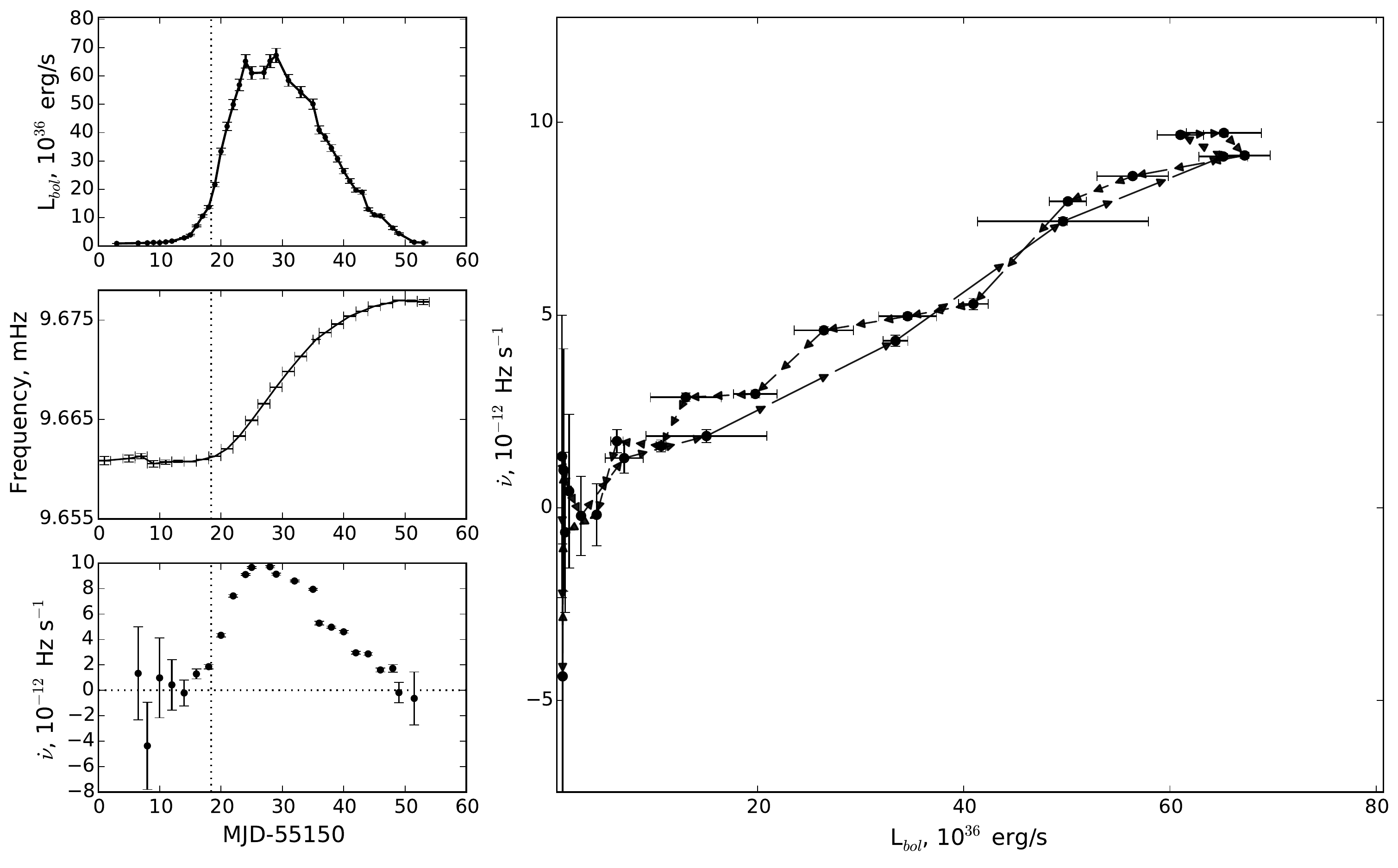}}
\put(250,270){\a0535}
\end{picture}
\center{\small\textbf{Fig.~\ref{observ}} --- Contd.}
\end{figure*}

\begin{figure*}

\begin{picture}(250,250)
\put(25,0){\includegraphics[width=2\columnwidth]{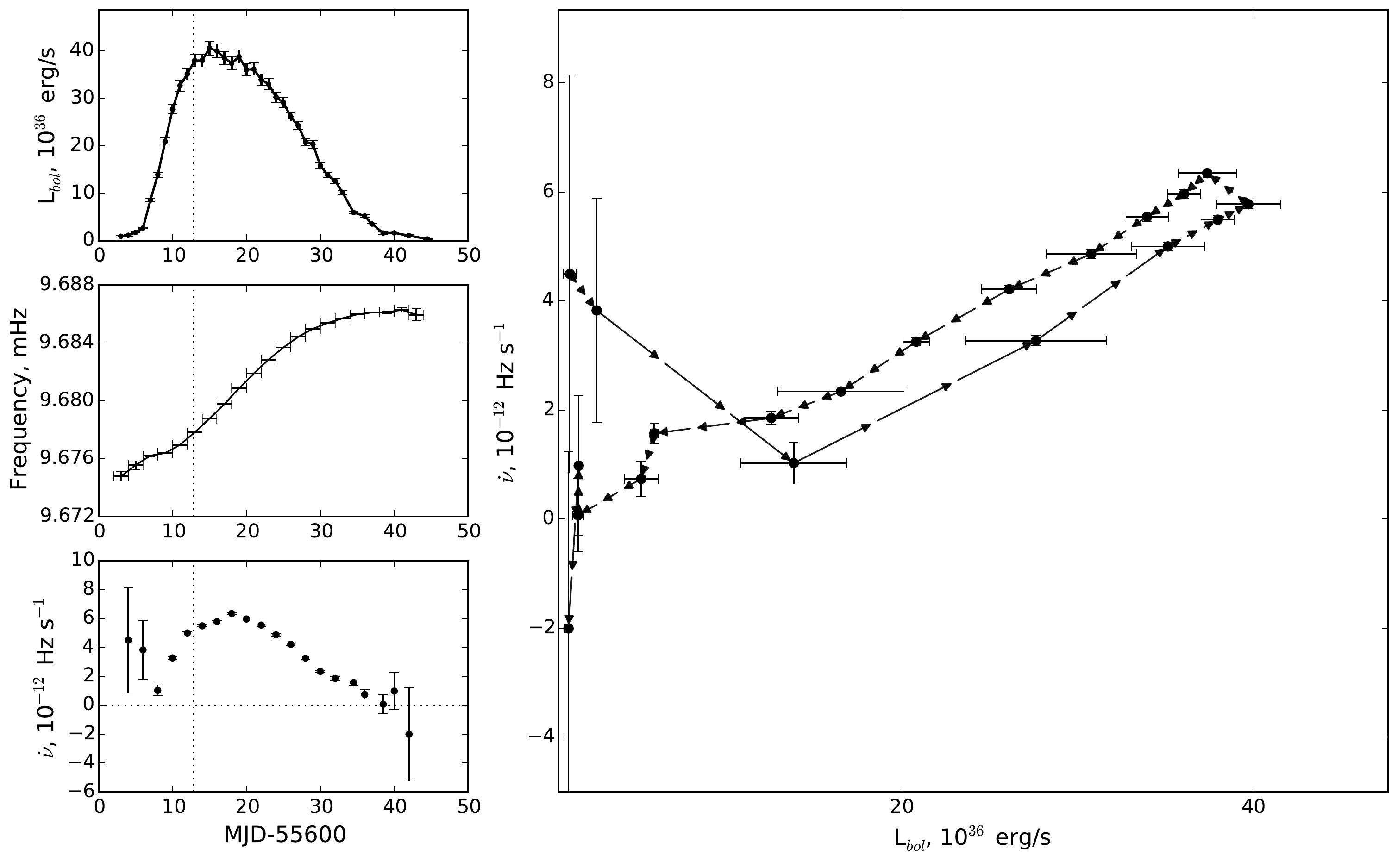}}
\put(250,270){\a0535}
\end{picture}
\center{\small\textbf{Fig.~\ref{observ}} --- Contd.}
\end{figure*}

\begin{table}
\caption{The dependence of the bolometric correction on the flux in  15-50 keV energy range, $F$ in erg s$^{-1}$cm$^{-2}$}\label{bol_cor}

\setlength{\extrarowheight}{5pt}

\begin{tabularx}{\columnwidth}{C|C}
\hline
Source& $k_{bol}$\\
\hline
\4u0115&$3.7-1.5\times 10^8 F$ \\
\v0332&$2.7+5.1\times 10^{7}F$\\
\rx0520& 2.5\\
\2s1553 &3.5\\
\xte1946 &1.9\\
\ks1947&2.2 \\
\g1008&$2.2-2.1\times 10^{7}F$\\
\a0535&$0.657F^{-0.0645}$\\

\hline
\end{tabularx}

\end{table}

\section*{SIZE OF THE NEUTRON STAR MAGNETOSPHERE}\label{simple_estimates}
During accretion the neutron star spin-up is determined
mainly by the angular momentum that is
transferred by  the accreting matter; $\dot{\nu}=N_{acc}/2\pi I$, where $I$ is the moment of inertia of the
neutron star taken to be $10^{45} $ g\, cm$^2$ and $N_{acc}=\dot{M}\sqrt{GM\xi r_A}$ during accretion through a disk \citep{frank85}. Substituting $\dot{M}= {{L_{bol}R_{NS}}\over{GM_{NS}}}$ yields a well-known dependence for disk accretion (in the case where the magnetosphere-disk interaction occurs only near the inner disk boundary):

\begin{equation}\label{nuacc}
\dot{\nu}=3.6\times 10^{-13}\xi^{1/2}L_{36}^{6/7}\mu_{30}^{2/7} (M_{NS})^{-3/7} R_{NS}^{6/7}I_{45}^{-1},    
\end{equation}
where $M_{NS}$ is expressed in units of $\msun$, and $R_{NS}$ is in  $10^6$ cm.

\begin{figure*}
\centering
\vbox{
\includegraphics[width=1.8\columnwidth]{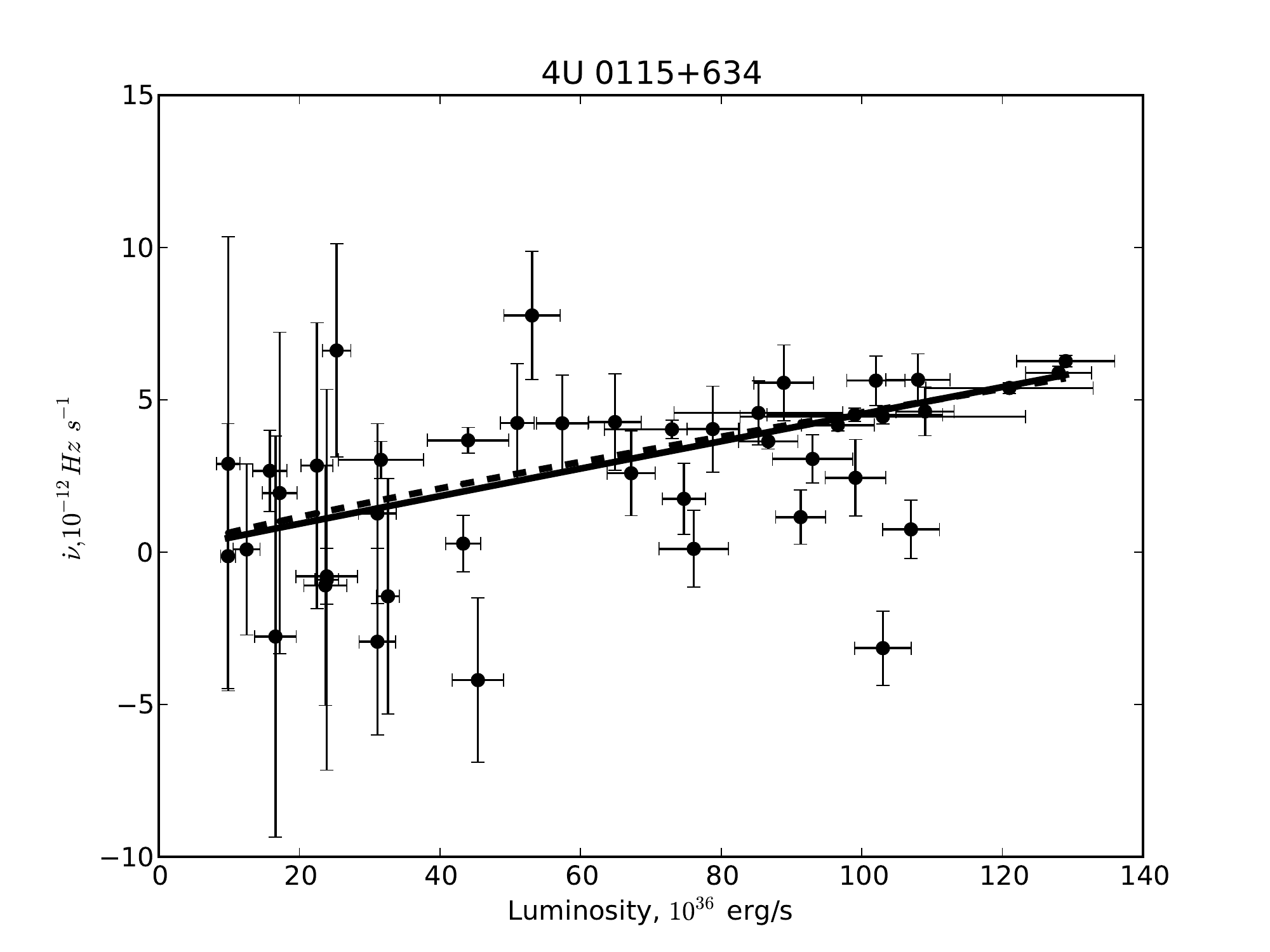}
\includegraphics[width=1.8\columnwidth]{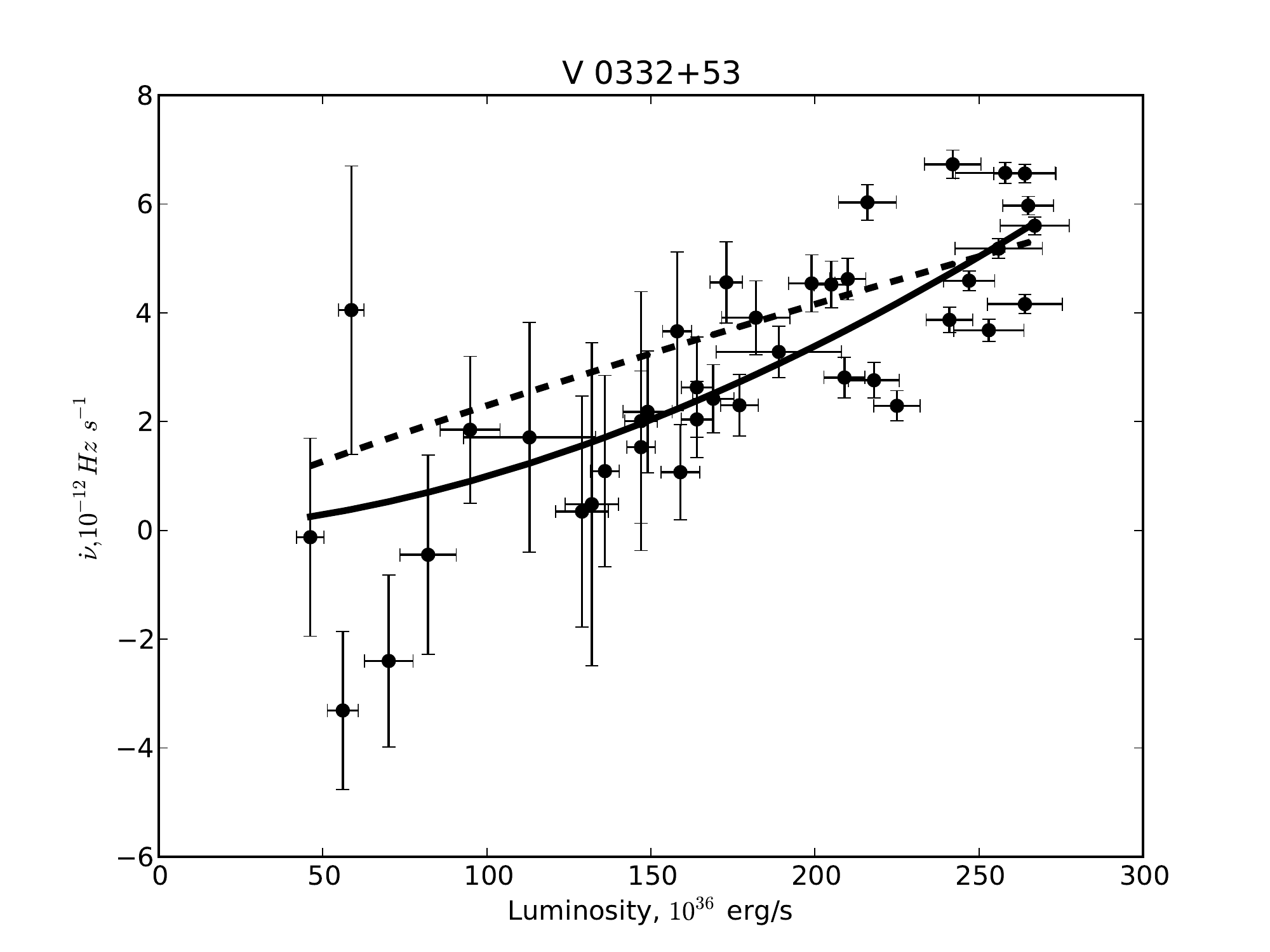}
}
\caption{The neutron star spin frequency derivative versus bolometric luminosity during type II outbursts. The solid line indicates
the fit to the data by a model $\dot{\nu}=A L^{\beta}$, where the exponent was a free parameter. The dashed line corresponds to the case where the exponent was fixed at $\beta=6/7$.}\label{dotnu_lum}
\end{figure*}

To determine how much the exponent in the observed dependence $\dot{\nu}(L_{bol})$ can differ from $6/7$, the
data shown in Fig. \ref{observ} were fitted by a simple power law $\dot{\nu}=AL^{\beta}$, where the exponent and the normalization
were free parameters. We fitted the data by the least squares method by taking into account the errors in the neutron star spin frequency derivative. All errors are specified at a level of one standard deviation (1$\sigma$). If several outbursts were recorded from the system,
then the data were fitted both for all outbursts simultaneously and for each individual outburst. The best fits to the data by the law 
$\dot{\nu}=AL^{\beta}$  for all outbursts of the source are presented in Table \ref{simple_pow} (second column) and Fig. \ref{dotnu_lum} (solid lines). A hysteresis behavior is observed in the systems \ks1947 and \v0332 during their outbursts. We fitted the increasing and decreasing
(in luminosity) branches separately and found that for \ks1947 on both branches the exponent is virtually the same; the increasing branch lies systematically above the decreasing one.  For \v0332 the increasing branch is below the decreasing one; $\beta_{incr}=1.4\pm0.2$ on the increasing branch and $\beta_{decr}=3.3\pm0.7$  on the decreasing one. It can be seen from Fig. \ref{observ} that a hysteresis behavior of $\dot{\nu}(L_{bol})$ during some outbursts of \a0535 and \g1008 is also observed.  In both systems the increasing
branch lies below the decreasing one; fitting the branches separately for \g1008 gives exponents that coincide, within the error limits, both between the branches of one outburst and from outburst to outburst. In \a0535 during the second outburst the exponent on the increasing branch slightly differs ($\beta_{incr}=1.5\pm0.1$) from that on the decreasing one ($\beta_{decr}=1.0\pm0.1$).

 It follows from Fig. \ref{dotnu_lum} and Table \ref{simple_pow} that most of the systems have an exponent close to the prediction $\dot{\nu}\sim L^{6/7}$. The system \v0332, for which the
exponent turned out to be about two, is an exception. Subsequently, to improve the statistical properties of the dependence $\dot{\nu}(L_{bol})$, the data for all outbursts from one system were fitted simultaneously. 

Fitting the data by Eq. (\ref{nuacc}) allows to estimate the size of the neutron star magnetosphere under assumption that the magnetosphere-accretion disk interaction occurs only near the inner disk boundary. The results are presented in Table 4, the second column ($\xi_{acc}$).
The best fit to the data by Eq. (\ref{nuacc}) is indicated in Fig. \ref{dotnu_lum} by the dashed line.

In order to estimate the size of the neutron star magnetosphere, we also used several analytical models that took into account the interaction of the magnetosphere with different parts of the accretion disk and contained an explicit dependence on the magnetospheric radius.

\begin{table*}
\caption{The exponent in $\dot{\nu}=AL^{\beta}$. Column 2 gives the fit for all outbursts from the source being investigated; column 3 gives the fit for the first outburst; column 4 gives the fit for the second outburst, if it exists. In the case where a hysteresis behavior of the dependence $\dot{\nu}(L_{bol})$ was observed during the outburst, the exponents on the increasing ($\beta_{incr}$) and decreasing ($\beta_{decr}$) branches are given separately}\label{simple_pow}
\setlength{\extrarowheight}{5pt}
\begin{tabularx}{2.0\columnwidth}{C|C|C|C|C|C}
\hline
Source & $\beta$&\multicolumn{2}{c|}{$\beta_{I out}$}& \multicolumn{2}{c}{$\beta_{II out}$}\\ 
\cline{3-6}
&&$\beta_{incr}$ & $\beta_{decr}$ & $\beta_{incr}$ & $\beta_{decr}$\\ 
\hline
\4u0115 &1.0(2)&\multicolumn{2}{c} {0.9(7)}&\multicolumn{2}{c}{0.9(2)}\\ 
\v0332 &1.8(5)&1.4(2)&3.3(7)&&\\ 
\rx0520& 0.7(2)&&&&\\ 
\2s1553 &0.6(2)&&&&\\ 
\xte1946 &0.9(2)&&&&\\ 
\ks1947&1.0(1)&1.0(1)&1.1(1)&&\\ 
\g1008&0.78(3)&0.9(1)&0.8(1)&\multicolumn{2}{c}{0.7(1)}\\ 
\a0535 &0.94(3)&\multicolumn{2}{c|}{0.91(5)}&1.5(1)&1.0(1)\\ 
\hline
\end{tabularx}
\end{table*}

\begin{table}
\caption{The size of the neutron star magnetosphere for several
models describing the dependence  $\dot{\nu}(L_{bol})$}\label{nuacc_ksi}

\setlength{\extrarowheight}{5pt}
\begin{tabularx}{\columnwidth}{l|C|C|C|C}

Source& $\xi_{acc}$&$\xi_{wang}$&$\xi_{li}$&$\xi_{parf}$\\
\hline
\4u0115 &0.1&0.1&0.04&0.3\\
\v0332 &0.02&0.01&0.01&0.2\\
\rx0520&0.8&0.6&0.3&--\\
\2s1553 &0.9&0.7&0.3&--\\
\xte1946 &2.2&1.8&1.1&--\\
\ks1947&2.2&1.7&0.9&--\\
\g1008&0.1&0.1&0.04&0.15\\
\a0535 &0.4&0.3&0.2&--\\

\hline
\end{tabularx}
\end{table}

\subsection*{The Models with a Neutron Star Magnetic Field Threading an Accretion Disk}

The first most complete analytical model where the stellar magnetosphere-accretion disk interaction was considered was published in \cite{ghosh79}. Subsequently, \cite{wang87} showed the internal inconsistency of this solution, while \cite{wang95} and \cite{li95} published
their calculations for self-consistent models. These models use the so-called fastness parameter $\omega=\Omega_s/\Omega_K(r_m)=(r_m/r_{co})^{3/2}$, where $\Omega_s$ is the stellar spin frequency, $\Omega_K$ is the Keplerian rotation frequency at radius $r_m$, and $r_{co}=(GM/\Omega_s^2)^{1/3}$ is the corotation radius of the star. Thus, the total torque is $n(\omega)N_{acc}$, where the dependence of $n(\omega)$ on the fastness parameter is determined by the configuration of the
interaction of the neutron star magnetic field with the accretion disk. 
Wang (1995) derived the formula:
\[
n(\omega)=(7/6-4/3\omega+1/9\omega^2)(1-\omega)^{-1}.
\]
The sizes of the neutron star magnetosphere $\xi_{wang}$ calculated from this formula are given in Table 4, the third column.

Li and Wang (1995) showed that the dependence can also be in the form:
\[
n(\omega)=1+{{20(1-31/16\omega)}\over{31(1-\omega)}}.
\]
The sizes of the neutron star magnetosphere $\xi_{li}$ derived in this model are given in Table 4, the fourth column. This approximation gives a magnetospheric size is about half that from the formula derived by Wang (1995). The estimate of the magnetospheric radius for V 0332+53 within the models from Wang (1995) and Li andWang (1995) is purely formal in nature, because, strictly speaking, they do not
describe the data (Fig. 4). 

Using the model from Li and Wang (1995), we estimated the magnetospheric size separately on the increasing and decreasing branches of the dependence $\dot{\nu}(L_{bol})$  for KS 1947+300, GROJ1008-57, and 1A 0535+26 during outbursts exhibiting hysteresis.
For KS 1947+300 $\xi\simeq0.7$ on the increasing branch  and $\xi\simeq1.0$ on the decreasing one. For \g1008 and \a0535 the magnetospheric size remains approximately the same from branch to branch and from outburst to outburst and is equal to the value
obtained by fitting all data for the source simultaneously.

\subsection*{The Model with Open Neutron Star Magnetic Field Lines}

In the model proposed by Parfrey et al. (2016) some fraction of the field lines of the neutron star magnetosphere $\zeta$ are assumed to become open as a result of its interaction with the accretion disk, which leads to an efficient neutron star spin-down. In our calculations we assumed that $\zeta=1$ \citep{parfrey17}. The total angular momentum being transferred to the star is defined by the formula for the
case where the magnetospheric radius $r_m$ is smaller than the corotation radius $r_{co}$. This leads to the following luminosity dependence of the neutron star spin frequency derivative:

\begin{equation}
\begin{multlined}
\dot{\nu}=3.6\times10^{-13}L_{36}^{6/7}\xi^{1/2}M_{NS}^{-3/7}R_{NS}^{6/7}\mu_{30}^{2/7}-\\
-7.2\times10^{-14}\xi^{-2}/P_{spin} L_{36}^{4/7} \mu_{30}^{6/7} M_{NS}^{-2/7} R_{NS}^{4/7}
\end{multlined}
\end{equation}

Ratio of the contributions from the spin-up and spin-down torques during an outburst shows that this model gives results different from the simple accretion model for three pulsars from our sample, 4U 0115+634, V 0332+53, and GRO J1008-57, as ratio lies within the ranges $\sim$1.4-3, 0.86-1.4, and 1.2-6, correspondingly. In the remaining cases, the spin-up torque due to the accretion of matter exceeds considerably the spin-down one, and this model gives
the same magnetospheric radius as does the model that takes into account only the accretion of matter (Eq. (2)). The fifth column in Table 4 gives the magnetospheric sizes $\xi_{parf}$ obtained within this model. For V 0332+53 we fitted the increasing and decreasing branches of the dependence $\dot{\nu}(L_{bol})$ separately. The inferred relativemagnetospheric radii coincide, within the error limits, with one another and with the value obtained by fitting both branches simultaneously.

\begin{figure*}
\centering
\vbox{
\includegraphics[width=1.8\columnwidth]{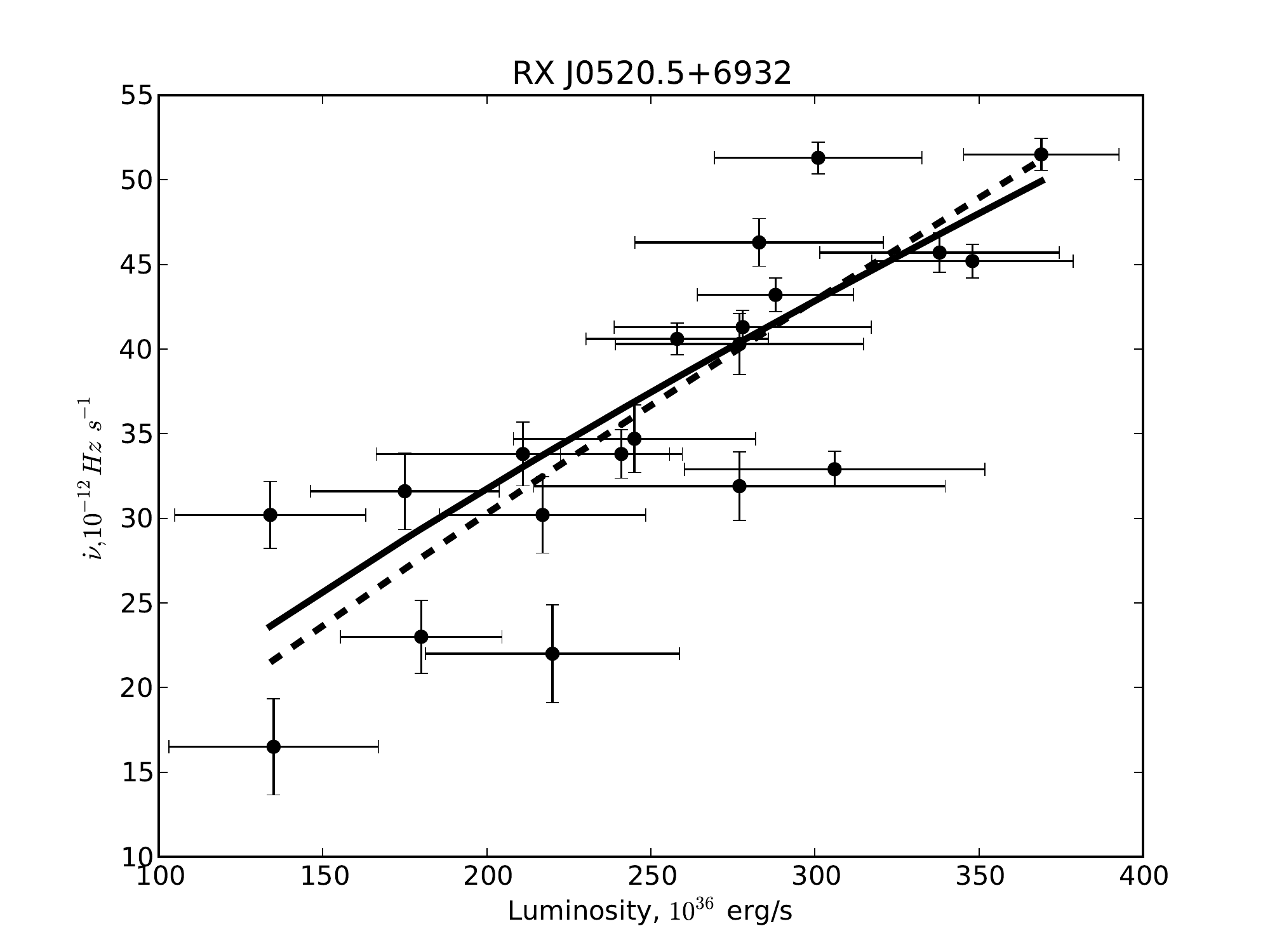}
\includegraphics[width=1.8\columnwidth]{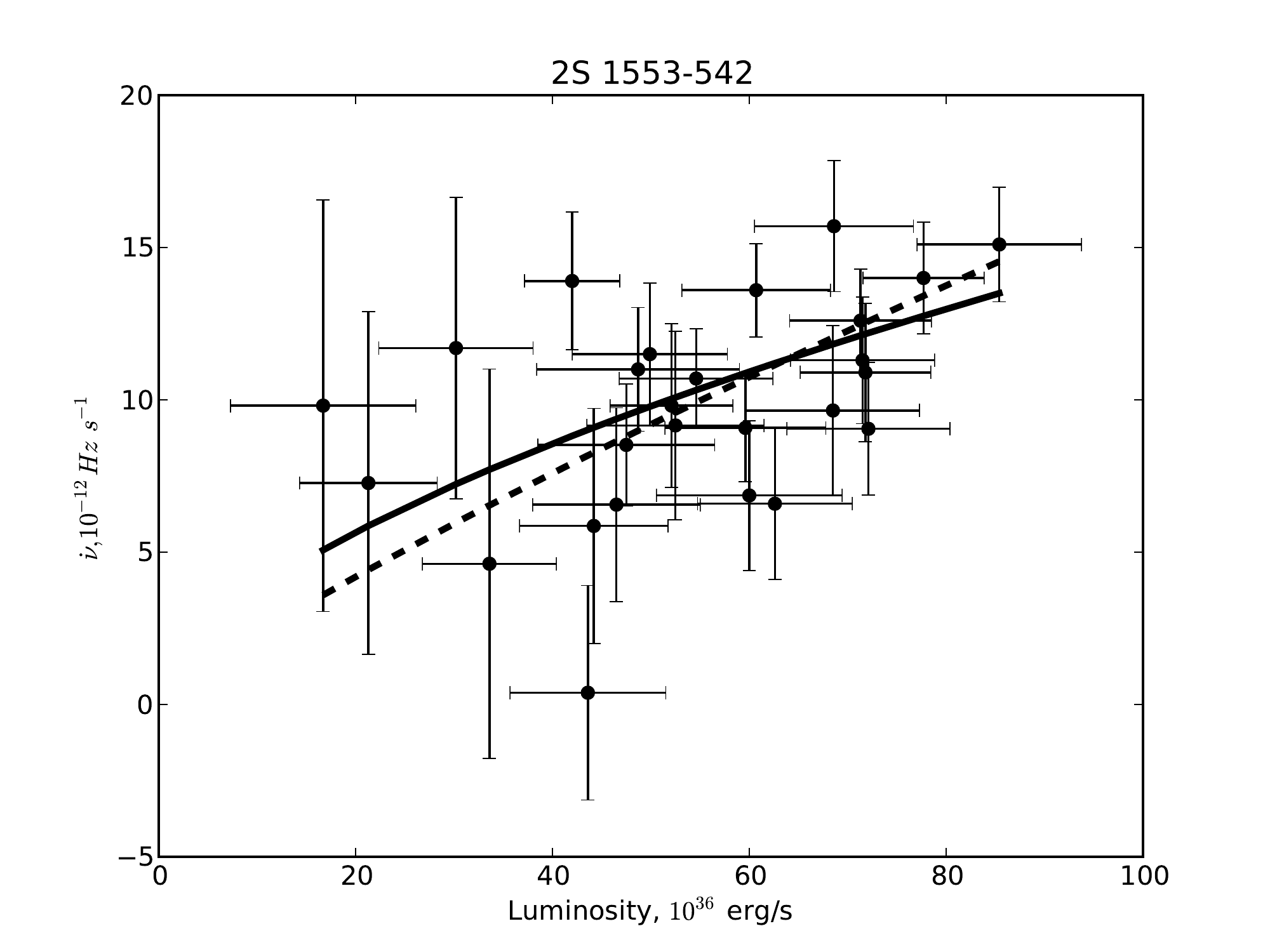}
}
\center{\small\textbf{Fig.~\ref{dotnu_lum}} --- Contd.}
\end{figure*}

\begin{figure*}
\centering
\vbox{

\includegraphics[width=1.8\columnwidth]{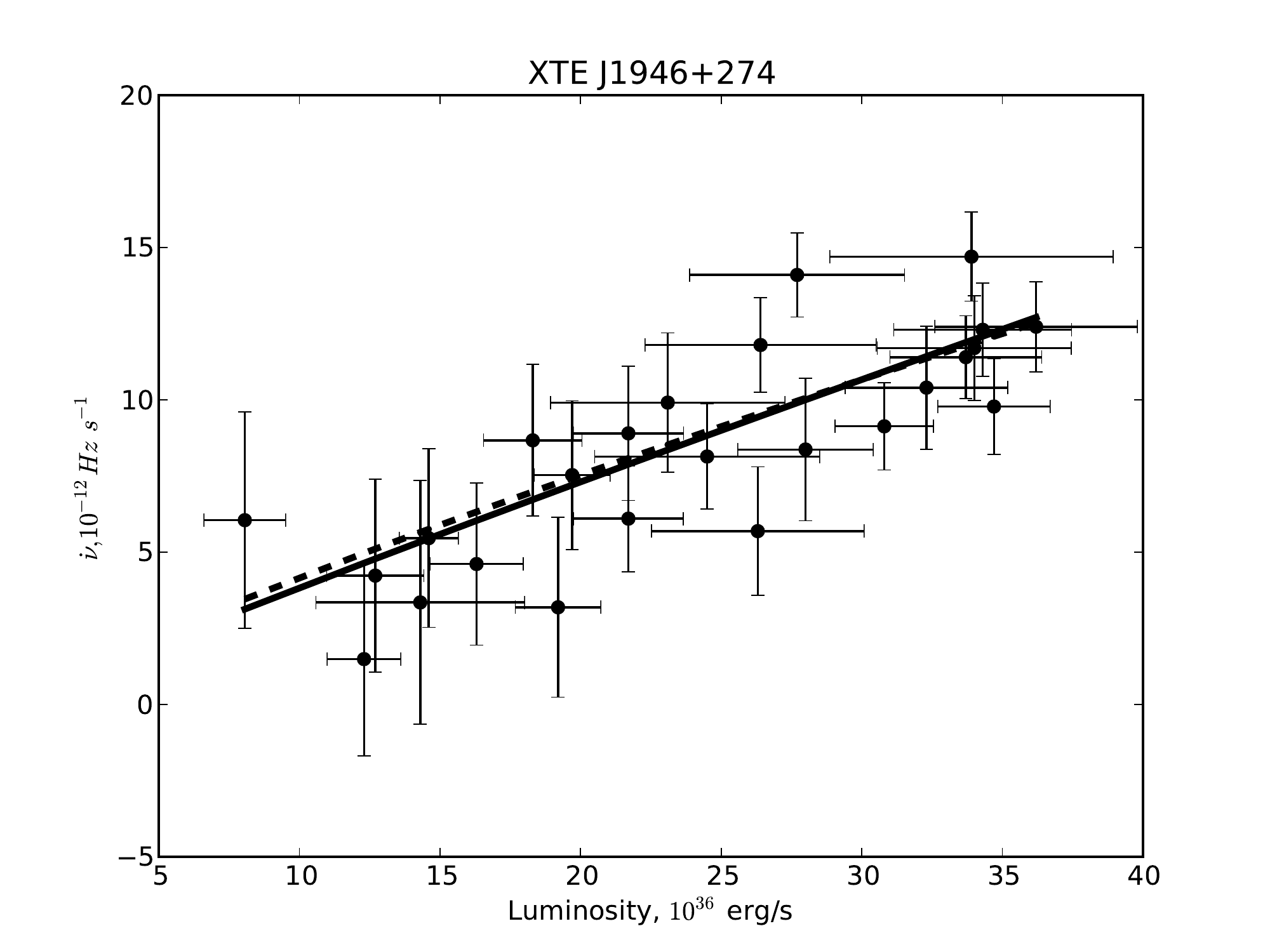}
\includegraphics[width=1.8\columnwidth]{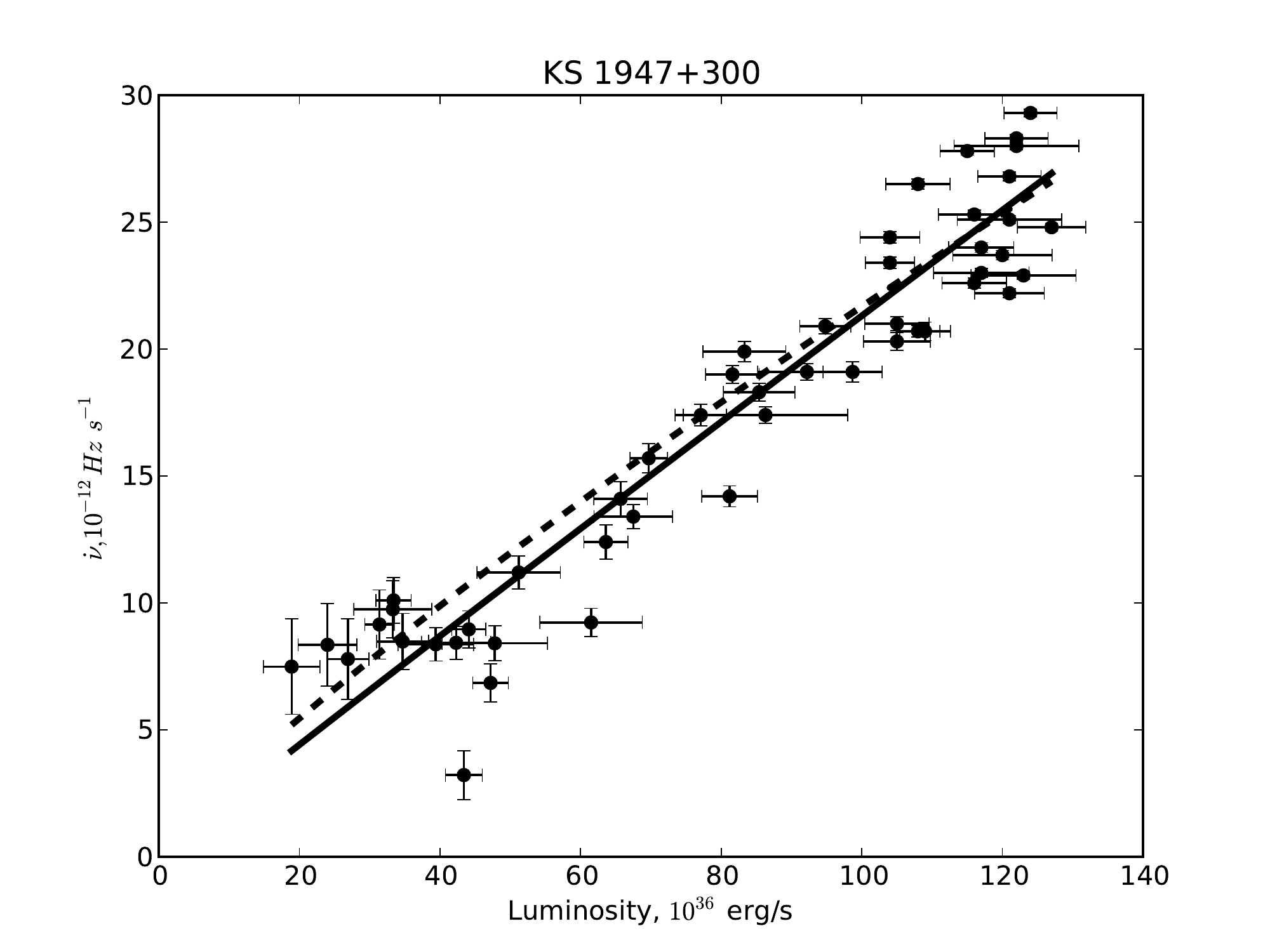}
}
\center{\small\textbf{Fig.~\ref{dotnu_lum}} --- Contd.}
\end{figure*}

\begin{figure*}
\centering
\vbox{
\includegraphics[width=1.8\columnwidth]{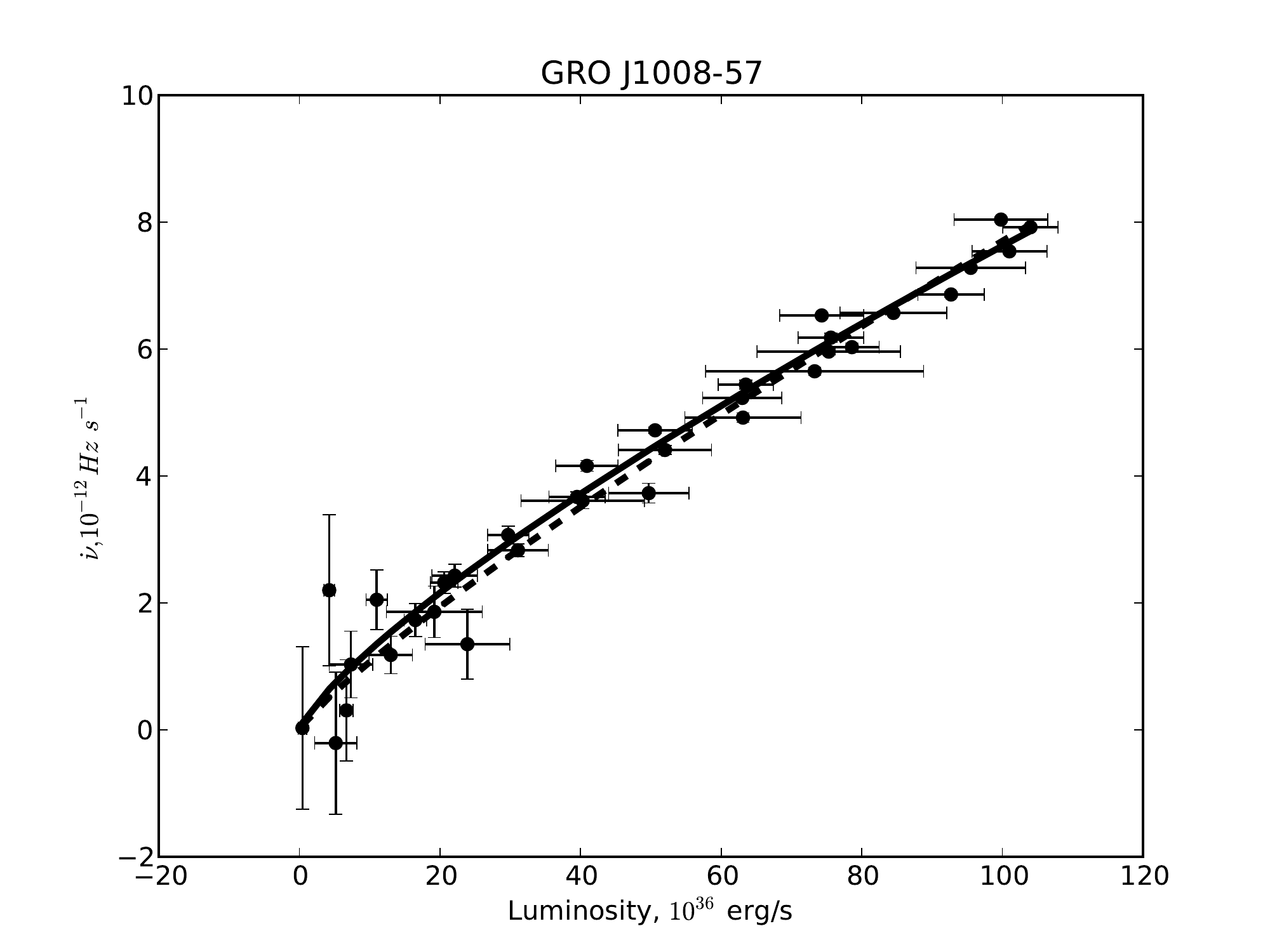}
\includegraphics[width=1.8\columnwidth]{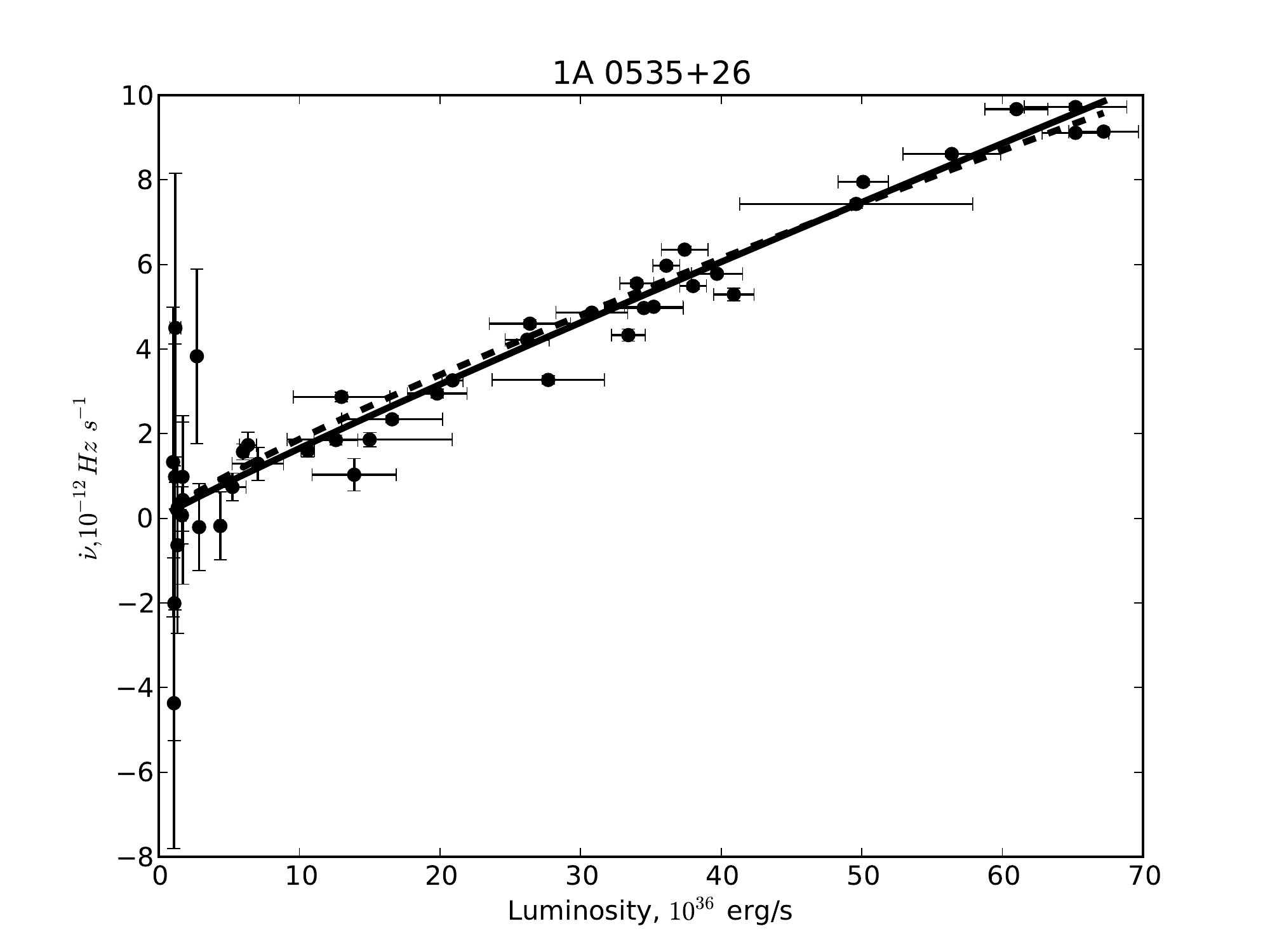}
}
\center{\small\textbf{Fig.~\ref{dotnu_lum}} --- Contd.}
\end{figure*}

\section*{ESTIMATING THE UNCERTAINTY IN THE MAGNETOSPHERIC RADIUS}

It can be seen from Eqs. (2)-(4) that $\dot{\nu}(L_{bol})$ depends on the neutron star mass and radius. Theoretical estimates predict that the radius of a neutron star with a mass of $1.4 \msun$ can be 14 km \citep{hebeler10,suleimanov11}, while most of the measured radii lie within the range 9.9-11.2 km \citep{ozel16}. It is also known from observations that the neutron star masses can change from 1 to 2.1$\msun$ for different systems \citep{nice05,falanga15}. To get an idea about the accuracy of the relative magnetospheric radius estimations, we analyzed the dependence of the magnetospheric radius on the uncertainty in the neutron star mass and radius and the distance to the system in the model from \cite{li95} for all systems (except for V 0332+53, because it does not
describe the observational data for this system) and the model from Parfrey et al. (2016) for the systems for which it is applicable. The neutron star mass was varied in the range $(1-2) \msun$; the neutron star radius was varied in the range $(1-1.5)\times 10^6$ cm. We
analyzed how the magnetospheric size changed with neutron star mass and radius at a fixed distance to the system and how it depended on the uncertainty in the distance to the system (see Table 1) at typical neutron star masses and radii, $M_{NS}=1.4\msun$ and $R_{NS}=10^6$ cm. For the model from Parfrey et al. (2016) the range of magnetospheric radii is given for the uncertainty both  in the distance to the system and in the neutron star mass and radius, because this dependence turned out to be insignificant. The inferred ranges of the neutron star magnetosphere radius are given in Table 5.

 It can be seen from Table 5 that even if the uncertainty in the distance to the system and the neutron star mass and radius are taken into account, the radius of the neutron star magnetosphere remainÙ within a narrow range, 0.1 - 0.3, for three systems,
4U 0115+634, V 0332+53, and GRO J1008-57. For the others the possible range of magnetospheric
radii is wide, from 0.1 to $\sim1$.

\begin{table}
\caption{The range of possible sizes of the neutron star magnetosphere when the uncertainty in the neutron star mass and radius and the uncertainty in the distance to the system are taken into account}\label{ksi_range}

\setlength{\extrarowheight}{5pt}
\begin{tabularx}{\columnwidth}{l|C|C|C}
\hline
Source&$\xi_{li}$, &$\xi_{li}$, fix.& $\xi_{parf}$\\
&fix. dist. & $M_{NS},R_{NS}$&\\
\hline
\4u0115&0.01-0.04&0.02-0.05&0.2-0.3\\
\v0332 &--&--&0.2-0.3\\
\rx0520&0.1-0.4&0.3&--\\
\2s1553 &0.1-0.5&0.2--0.8&--\\
\xte1946 &0.2-1.3&0.6-1.1&--\\
\ks1947&0.1-1&0.5-1.&--\\
\g1008&0.01-0.03&0.03-0.05&0.1-0.2\\
\a0535&0.03-0.2&0.1-0.7&--\\
\hline
\end{tabularx}
\end{table}

\section*{CONCLUSIONS}

We carried out a systematic study of the spin frequency derivative of X-ray pulsars as a function of the bolometric luminosity, $\dot{\nu}(L_{bol})$,  or the mass accretion rate onto them. This dependence is an important test for the theories of angular momentum
transfer during accretion and the structure of the inner accretion disk. It is important to note that a proper construction of the experimental dependence $\dot{\nu}(L_{bol})$ requires additional information to estimate the bolometric luminosity of the source and to correct
the pulsation frequency for the orbital motion, which was done in this paper. At present, there is a large set of observational data from the Fermi/GBM telescope, which has continuously monitored X-ray pulsars and traced the history of their pulsation frequency changes since 2008, and data from the Swift/BAT telescope, which has constructed continuous light curves of sources in the hard X-ray band (15-50 keV)
since 2005. Based on these observations, we were able to construct the dependence  $\dot{\nu}(L_{bol})$ for eight X-ray pulsars in Be systems during type II outbursts, which allowed this dependence to be investigated in a wide range of mass accretion rates. We corrected
the Fermi/GBM data for the orbital motion by ourselves for in the systems XTE J1946+274 and GRO J1008-57 using their orbital parameters from the literature and for the source KS 1947+300 using improved orbital parameters.

To a first approximation, the neutron star spins up during outbursts due to the transfer of angular momentum by the accreting matter. In this case, the dependence $\dot{\nu}(L_{bol})$ is a power law with an exponent of 6/7. \cite{ghosh79,wang95,li95} showed that the interaction of the neutron star magnetosphere with the accretion disk also causes the dependence $\dot{\nu}(L_{bol})$ to be a power law
with an exponent of 6/7. Our fitting the experimental data by a simple power law $\dot{\nu}\sim L_{bol}^{\beta}$ gives an exponent close to 6/7 for most systems, with the exception of V 0332+53 for which $\beta=1.8\pm0.5$. For several systems we obtained a value of $\beta$ different from 6/7 within the 1$\sigma$ error limits (KS 1947+300, 1A 0535+26, 2S 1553-542, and GRO J1008-57). 
However, the errors in $\dot{\nu}$ shown in Figs. 3 and 4 are statistical in nature and may be underestimated, because the systematic measurement error (related, for example, to the inaccuracy of determining the orbital parameters) is disregarded, which, in
turn, leads to an underestimation of the error in the slope in the power-law dependence. Thus, the available data do not allow one to distinguish the dependences from \cite{ghosh92} for a radiation-dominated disk and a two-temperature,  optically thin in the vertical direction gas-dominated disk, where the electrons cool through their bremsstrahlung (see the Introduction), but allow to exclude the model of a two-temperature, optically thin in the vertical direction gas-dominated disk, where the electrons cool through Comptonization of soft photons from the external source.

A hysteresis in neutron star spin-up behavior during an outburst was detected in the systems KS 1947+300, GRO J1008-57, and 1A 0535+26 and confirmed in the system V 0332+53. In V 0332+53 the spin-up on the increasing branch is systematically higher than
that on the decreasing one. At the same time, in KS 1947+300, GRO J1008-57, and 1A 0535+26 the  spin-up on the increasing branch is systematically lower than that on the decreasing one. For these systems we fitted the increasing and decreasing branches of the dependence $\dot{\nu}(L_{bol})$ separately and found that in KS 1947+300 and GRO J1008-57 the exponent barely changes and is equal to the exponent obtained by fitting both branches simultaneously. In 1A 0535+26 $\beta_{incr}=1.5\pm0.1$ on the increasing
branch and $\beta_{decr}=1.0\pm0.1$  on the decreasing one. However, this difference may be related to the systematic
deviations due to insufficient coverage of the outburst onset by the observations. In V 0332+53 $\beta_{incr}=1.4\pm0.2$  on the increasing branch and $\beta_{decr}=3.3\pm0.7$  on the decreasing one. 

We estimated the size of the neutron star magnetosphere in the approximation of the simple accretion model (Eq. (2)) and the models from Wang (1995) and Li and Wang (1995). Owing to the independent measurements of the magnetic field strength from a cyclotron absorption line in their energy spectra, the size of the neutron star magnetosphere was the only free parameter
in the models under consideration (the absence of any deviation of the magnetic field structure from a dipole configuration was demonstrated for several pulsars by Tsygankov et al. (2016a) and Lutovinov et al. (2017)). The magnetospheric size for the model from Wang (1995) is close to the values within the simple accretion model, while the magnetospheric size is approximately half as large for the model from Li and Wang (1995). It is important to note that the magnetospheric radius $r_{m}<r_{A}$ for most sources, but for two objects, XTE J1946+274 and KS 1947+300, it turned out to be about $2r_A$. \citep{chashkina17} showed that the magnetospheric size could reach  $2r_A$ for a radiation-dominated disk with a large geometrical thickness. However, at the luminosities demonstrated by both systems $10^{37}-10^{38}$ erg s$^{-1}$ and magnetic moments of $\sim (0.7-2)\times10^{30}$ G cm$^3$ these calculations predict an accretion disk thickness $H/R\leq0.03$, i.e., the disk must be a standard one, and the size of the neutron star magnetosphere  $\sim$0.3-0.5. The behavior of the dependence $\dot{\nu}(L_{bol})$ during an outburst in V 0332+53 suggests that the models from Wang (1995) and Li and Wang (1995) do not work for it; for its investigation we used the model from \cite{parfrey16}, which can also be
applied for 4U 0115+634 and GRO J1008-57. For the systems V 0332+53, KS 1947+300, GRO J1008-57, and 1A 0535+26 we estimated the magnetospheric size on the increasing and decreasing branches of the dependence $\dot{\nu}(L_{bol})$ separately and found that in V 0332+53, GRO J1008-57, and 1A 0535+26 the magnetospheric size does not change from branch to branch, while in KS 1947+300
there is some difference: the magnetospheric size on the increasing branch is smaller than that on the decreasing one by a factor of 1.4.

In conclusion, it should be noted that the used models were constructed in the approximation of an aligned rotator. This is not a quite correct assumption, because pulsations are observed from the neutron stars. So far there is no complete analytical model that takes this effect into account. In the literature there are studies where the individual components of accretion onto an oblique rotator are considered. In particular, \cite{perna06} considered the simplest model of an oblique rotator where the pulsar is in the propeller regime and part of the matter does not leave the system but returns back into the disk when interacting with the neutron star magnetosphere. \cite{kulkarni13} performed numerical simulations of matter accretion onto a strongly magnetized misaligned rotator (the angle between the rotation and magnetic axes was $\theta<30^{\circ}$) and the dependence of the radius of the neutron star magnetosphere on the accretion rate $\xi \sim 0.77 (\mu^2/\dot{M})^{-0.086}$ was derived. For all of the pulsars from our sample $\xi$ derived from this formula lies within the range $0.1-0.2$.

\section*{ACKNOWLEDGMENTS}

This work was supported by the Russian Science Foundation (grant no. 14-12-01287). We used the data from the Fermi/GBM observatory, the Swift Science Data Center at the University of Leicester,  and the data from the MAXI telescope provided by RIKEN, JAXA, and the MAXI team. We wish to give special thanks to P. Jenke for his help in working with the Fermi/GBM data and A. Beloborodovy for the discussion of results and useful comments.

\end{document}